\documentclass[11pt]{article}

\usepackage{setspace}
\usepackage{fullpage}
\usepackage{enumitem}
\usepackage[page]{appendix}
\usepackage{amssymb}
\usepackage{mathrsfs} 
\usepackage{natbib}
\usepackage{algorithm}
\usepackage{algorithmic}
\usepackage{hyperref}
\usepackage{comment}
\usepackage{multirow}
\usepackage{amsthm}
\usepackage{mathtools}
\usepackage{epstopdf,xcolor}
\usepackage{bbm}
\usepackage{dsfont}
\usepackage{pifont}
\usepackage{subcaption}
\usepackage{tikz, subcaption}
\usetikzlibrary{shapes, arrows, positioning}
\usepackage{authblk}
\usepackage{listings}
\usepackage{framed}

\definecolor{royalblue}{rgb}{0.25, 0.41, 0.88}

\newtheorem{theorem}{Theorem}
\newtheorem{lemma}{Lemma}

\newtheorem{proposition}{Proposition}
\newtheorem{corollary}{Corollary}
\newtheorem{definition}{Definition}
\newtheorem{example}{Example}
\newtheorem{remark}{Remark}

\newtheorem{assumption}{Assumption}

\newcommand{\E}{\mathbb{E}}

\newcommand{\Hc}{\mathcal{H}}

\newcommand{\hath}{\hat h_{\lambda,t}^\text{IF}}
\newcommand{\starh}{h^*_{\lambda,t}}

\newcommand{\independent}{\perp\mkern-9.5mu\perp}

\usepackage[normalem]{ulem}
\newcommand{\stkout}[1]{\ifmmode\text{\sout{\ensuremath{#1}}}\else\sout{#1}\fi}

\usepackage{hyperref}
\hypersetup{
    colorlinks= true,
    linkcolor= black,
    filecolor= cyan,      
    citecolor= royalblue
}

\DeclareMathOperator*{\argmin}{arg\,min}

\title{\bf \LARGE Debiased Ill-Posed Regression\\~\\}  

\author[1]{AmirEmad Ghassami\thanks{Correspondance Email: \texttt{ghassami@bu.edu}}}
\author[2]{James M. Robins}
\author[3]{Andrea Rotnitzky}
\affil[1]{Department of Mathematics and Statistics, Boston University}
\affil[2]{Harvard T.H. Chan School of Public Health, Harvard University}
\affil[3]{Department of Biostatistics, University of Washington}

\date{May 27, 2025}

\begin{document}

\maketitle

\begin{abstract}
In various statistical settings, the goal is to estimate a function which is restricted by the statistical model only through a conditional moment restriction. Prominent examples include the nonparametric instrumental variable framework for estimating the structural function of the outcome variable, and the proximal causal inference framework for estimating the bridge functions. A common strategy in the literature is to find the minimizer of the projected mean squared error. However, this approach can be sensitive to misspecification or slow convergence rate of the estimators of the involved nuisance components. In this work, we propose a debiased estimation strategy based on the influence function of a modification of the projected error and demonstrate its finite-sample convergence rate. Our proposed estimator possesses a second-order bias with respect to the involved nuisance functions and a desirable robustness property with respect to the misspecification of one of the nuisance functions. The proposed estimator involves a hyper-parameter, for which the optimal value depends on potentially unknown features of the underlying data-generating process. Hence, we further propose a hyper-parameter selection approach based on cross-validation and derive an error bound for the resulting estimator. This analysis highlights the potential rate loss due to hyper-parameter selection and underscore the importance and advantages of incorporating debiasing in this setting. We also study the application of our approach to the estimation of regular parameters in a specific parameter class, which are linear functionals of the solutions to the conditional moment restrictions and provide sufficient conditions for achieving root-n consistency using our debiased estimator.\\

\end{abstract}

\section{Introduction}
\label{sec:intro}

In many applications in statistics, particularly in causal inference, given a random sample drawn from the unknown distribution of a random vector $V,$ we are interested in estimating a function $h$ satisfying a conditional moment restriction of the form 
\begin{equation}
\label{eq:cmrNN}
\E\left[g_1(V)h(V_h)-g_0(V)\mid V_q\right]=0,
\end{equation}
where $V_{h},V_{q}\subseteq V$, and $g_{0}$ and $g_{1}$ are known functions. If $V_{h}=V_{q}$, Equation \eqref{eq:cmrNN} can be reformulated as a standard regression problem. Yet, we consider the more challenging case where $V_{h}\neq V_{q}$. In this case, Equation \eqref{eq:cmrNN} is an ill-posed integral equation \citep{kress2013linear}. We refer to the task of estimating a function $h$ that solves this equation as \emph{ill-posed regression}. Two of the most prominent applications of ill-posed regression in the filed of causal inference are found in instrumental variable (IV) and in proximal causal inference methodologies. In both cases, the goal is to perform causal inference in settings where unobserved confounders in the treatment-outcome relationship cannot be ruled out. 

IV methods require the existence of an auxiliary variable, called an instrumental variable, which only affects the outcome variable through the treatment variable and does not share an unobserved confounder with the outcome variable. A particularly popular IV-based model is the nonparametric IV (NPIV) model \citep{newey2003instrumental, hall2005nonparametric, darolles2011nonparametric, florens2011identification,chen2012estimation}. This model formalizes the aforementioned IV conditions as follows. Let $W\in \mathcal{W}$ and $Y\in \mathcal{Y}$ be the treatment and the outcome variables, respectively. The NPIV model assumes that the outcome variable $Y$ is generated from a structural function of the treatment variable, $h_{0}(W)$, with additive noise. That is, the outcome is generated as 
\[
Y=h_{0}(W)+\epsilon ,
\]
where $\epsilon$ is a mean-zero error term. Interest is in estimation of the unknown function $h_{0}$ based on random sample from the target population, when $W$ is not assumed to be exogenous, and hence $\mathbb{E}[\epsilon \mid W]$ is not assumed to be equal to zero. Therefore, $\mathbb{E}[Y\mid W=\cdot ]$ is not necessarily equal to the structural function $h_{0}\left( \cdot \right) $. Instead, in the NPIV model, we assume existence of an instrumental variable $Z\in \mathcal{Z}$, for which we have $\mathbb{E}[\epsilon \mid Z]=0$. That is, $h_{0}$ satisfies the conditional moment restriction 
\begin{equation*}
\mathbb{E}[h(W)-Y\mid Z]=0.  
\end{equation*}
Estimation of $h_{0}$ is a special case of the introduced ill-posed regression with $V_{q}=\{Z\}$, $V_{h}=\{W\}$, $g_{1}(V)=1$, and $g_{0}(V)=Y$. In words, this conditional moment restriction requires that after projecting onto the space of functions of $Z$, random variables $Y$ and $h_{0}(W)$ have the same projection. This requirement implies that the IV affects the outcome variable only through the treatment variable. 

Proximal causal inference methods \citep{miao2018identifying, tchetgen2020introduction, cui2023semiparametric} allow for the presence of unobserved confounders but require access to two proxy variables of the confounder, which satisfy certain statistical conditions. Consider, for example, the goal of estimating the mean of an  outcome of interest in the counterfactual world in which all subjects in the population receive the same treatment, based on a random sample of $\left( A,X,Y,Z,W\right) $, where $A=1$ if the subject received the treatment of interest and $A=0$ otherwise,  $X$ is a vector of observed baseline covariates, $Z$ and $W$ are the treatment and outcome proxy variables (defined in Section \ref{sec:proximal}), and $Y$ is the outcome variable. Under the proximal causal inference assumptions (detailed in Section \ref{sec:proximal}), the counterfactual mean under treatment $A=a$, for $a\in\{0,1\}$, is equal to $\mathbb{E}\left[ h_{0}\left( W,X\right) \right] $ where $h_{0}$  solves the conditional moment equation 
\[
\mathbb{E}[I(A=a)h(W,X)-I(A=a)Y\mid Z,A,X]=0.
\]
Estimation of $h_{0}$ is another special case of ill-posed regression with $V_{q}=\{Z,A,X\}$, $V_{h}=\{W,X\}$, $g_{1}(V)=I(A=a)$, and $g_{0}(V)=I(A=a)Y$. Under slightly different assumptions, \cite{cui2023semiparametric}, showed that the counterfactual mean is equal to the mean of $YI(A=a)q_{0}\left(Z,X\right) $, where $q_{0}(Z,X)$ solves the conditional moment equations  
\[
\E\left[I(A=a)q(Z,A,X)-1\mid W,X\right]=0,
\]
which is another special case of ill-posed regression with $V_{q}=\{W,X\}$, $V_{h}=\{Z,A,X\}$, $g_{1}(V)=I(A=a)$, and $g_{0}(V)=1$. 

The preceding examples illustrate the important role of ill-posed regression in causal inference, and motivates the development of solutions for this task. For a function $h$, define $\mathbb{E}[\mathbb{E}[g_{1}(V)h(V_{h})-g_{0}(V)\mid V_{q}]^{2}]$ as the projected error of $h$. Because the solution $h_{0}$ to Equation \eqref{eq:cmrNN} minimizes the projected error over all functions $h$, a common approach in the literature for ill-posed regression is to estimate $h_{0}$ with a minimizer of an estimate of (a regularized version of) the projected error. Therefore, the more precise the estimator of the projected error is, the better the quality of the estimated structural function, and hence, improving the estimator should be considered. Formalizing this idea is one of the main goals in this work.

In this paper, we propose a novel strategy for ill-posed regression which is based on minimizing a \emph{debiased} estimator of the projection error. We perform the debiasing step using techniques from the modern semiparametric theory, specifically, by leveraging the influence function of the projected error. The technique of influence function-based debiasing has recently gained significant attention in the literature of causal inference and missing data \citep{laan2003unified, van2011targeted, chernozhukov2018double}, starting from the seminal work of \cite{robins1994estimation}. For the instrumental variable setting, compared to existing work, our proposed method requires the estimation of an extra nuisance function, which is used for debiasing the original estimator of the projected error. We demonstrate that the proposed estimator will have a second-order bias with respect to the nuisance functions and possesses a desirable robustness property with respect to misspecification of one of its nuisance functions. As the first main result of this work, we provide finite-sample bounds on the convergence rate of our proposed estimator. The proposed estimator involves a hyper-parameter, for which the optimal value depends on certain functionals of the underlying unknown data generating process. Hence, we further propose a hyper-parameter tuning approach based on cross-validation. While there has been work in the literature studying whether or not hyper-parameter tuning has a detrimental effect on the convergence rate in standard non-parametric regression settings (e.g., \citep{vaart2006oracle, van2003unified}), to the best of our knowledge, the impact of hyper-parameter tuning on ill-posed regression is not well-understood. As our second main result, we derive finite-sample convergence bounds on the error of the cross-validated estimator. Our bounds point to the potential rate loss due to not knowing the optimal value of the hyper-parameter, and emphasize the importance of using the debiasing strategy in this setting.

In a recent work, \cite{ghassami2022minimax} considered a class of pathwise differentiable parameters $\psi _{0}$ with influence function of the form 
\[
IF_{\psi_{0}}(V)=s_{1}(V)q_{0}(V_{q})h_{0}(V_{h})+s_{2}(V)q_{0}(V_{q})+s_{3}(V)h_{0}(V_{h})+s_{4}(V)-\psi _{0},
\]
where $s_{1}$, $s_{2}$, $s_{3}$, and $s_{4}$ are known functions, i.e., they do not depend on the data generating law, and $h_{0}$ and $q_{0}$ are unknown functions. This is a generalization of the Robins' parameter class \citep{robins2008higher} to the case where $V_{h}\neq V_{q}$. It is shown in \citep{ghassami2022minimax} that the unknown functions $h_{0}$ and $q_{0}$ solve equations of the form \eqref{eq:cmrNN} for suitable choices of $g_{0}$ and $g_{1}.$  As will be explained in Section \ref{sec:proximal} and discussed in \citep{ghassami2022minimax}, the aforementioned counterfactual mean under the proximal causal inference assumptions is a parameter with influence function of the form as in the last display. The authors of \citep{ghassami2022minimax,kallus2021causal} proposed an adversarial learning approach for estimating $h_{0}$ and $q_{0}$, and studied the asymptotic properties of the debiased (influence function-based) estimator for $\psi _{0}$. Here, we apply our ill-posed regression strategy to the setting of \cite{ghassami2022minimax}, and extend the debiasing point of view by replacing the adversarial approach to the estimation of the nuisance functions $h_{0}$ and $q_{0}$ with estimators that minimize a debiased (influence function-based) estimator of the projected error. Hence, we propose an estimation strategy with two layers of debiasing. We demonstrate the gained benefit from this extra added debiasing layer in terms of the conditions needed for obtaining a consistent and asymptotically normal estimator for $\psi _{0}$. Importantly, we show that without using the debiasing strategy, achieving asymptotic normality may not be feasible.

The rest of the paper is organized as follows. In Section \ref{sec:desc}, we introduce the model that we study in this work. We present our debiased ill-posed regression approach in Section \ref{sec:DIVreg}, and analyze its bias structure. We present the finite-sample convergence rate of our proposed estimator in Section \ref{sec:mainres}. We describe the hyper-parameter tuning approach and its convergence analysis in Section \ref{sec:HPselect}. We study the application of our proposed ill-posed regression approach in the parameter class of \cite{ghassami2022minimax} and illustrate it with the estimation of the counterfactual mean under the proximal causal inference framework in Section \ref{sec:linfuncal}. Our concluding remarks are presented in Section~\ref{sec:conc}.

\section{The Integral Equation}
\label{sec:desc}

Given a law $P$ on the sample space of $V,$ we let $P_{V_{h}}$ and $P_{V_{q}}$ be the corresponding marginal laws on the sample spaces of $V_{h}\subset V$ and $V_{q}\subset V$, respectively. We let $L^{2}(P_{V_{h}})$ (resp. $L^{2}(P_{V_{q}})$) be the set of measurable functions on the range of $V_{h}$ (resp. $V_{q}$) with finite second moment under $P_{V_{h}}$ (resp. $P_{V_{q}}$). We denote the law of $V$ by $P_{0}$. Let $T:L^{2}(P_{0,V_{h}})\rightarrow L^{2}(P_{0,V_{q}})$ be the linear, bounded, operator $(Th)(V_{q})=\mathbb{E}[g_{1}(V)h(V_{h})\mid V_{q}]$, where $g_{1}$ is a known function and throughout, $\mathbb{E}$ stands for expectation under $P_{0}$. Moreover, let $r_{0}\in L^{2}(P_{0,V_{q}})$ be defined as $r_{0}(V_{q})=\mathbb{E}[g_{0}(V)\mid V_{q}]$, where $g_{0}$ is a known function. The conditional moment restriction in Equation \eqref{eq:cmrNN} can be written using the operator notation as 
\begin{equation}
\label{eq:IP1}
Th=r_{0}.  
\end{equation}
Throughout we assume that a solution to this equation exists, i.e., we make the following assumption.
\begin{assumption}
\label{assm:exists} 
$r_{0}\in \mathcal{R}(T)$, where $\mathcal{R}(T)$ is the range of the operator $T$.
\end{assumption}

When the operator $T$ is compact (see, e.g., \citep{kress2013linear}), a situation that can only arise when $V_{h}$ and $V_{q}$ do not share common components, a sufficient condition for $r_{0}\in \mathcal{R}(T)$  is provided by the celebrated Picard's theorem, which we review next. To proceed, we remind the reader of an important property of compact linear operators. For any compact linear operator $T:\mathcal{E}\rightarrow \mathcal{E}'$ between Hilbert spaces $\mathcal{E}$ and $\mathcal{E}'$, let $(\sigma _{i})_{i=1}^{\infty }$ denote the sequence of, decreasing to 0, square roots of the eigenvalues of the self-adjoint compact operator $T^{\ast }T,$ where $T^{\ast }$ denotes the adjoint of $T$. Then, there exist orthonormal sequences $(\varphi _{i})_{i=1}^{\infty }$ and $(\psi_{i})_{i=1}^{\infty }$ in $\mathcal{E}$ and $\mathcal{E}'$, respectively, such that for all $i$:  
\[
T\varphi _{i}=\sigma _{i}\psi _{i},~~~~~T^{\ast }\psi _{i}=\sigma
_{i}\varphi _{i}.
\]
The sequence $(\sigma _{i},\varphi _{i},\psi _{i})_{i=1}^{\infty }$ is called a singular system of $T$.

\begin{theorem}[Picard's Theorem]
Let $T:\mathcal{E}\rightarrow \mathcal{E}'$ be a compact linear operator with singular system $(\sigma _{i},\varphi _{i},\psi _{i})_{i=1}^{\infty }$. A solution to the integral equation $Th=r_{0}$ is identifiable if $r_{0}$ belongs to $\mathcal{N}(T^{\ast })^{\perp }=\overline{\mathcal{R}(T)}$ and satisfies 
\[
\sum_{i=1}^{\infty }\frac{1}{\sigma _{i}^{2}}|\langle r_{0},\psi _{i}\rangle
|^{2}<\infty ,
\]
where $\mathcal{N}(T^{\ast })$ is the null space of $T^{\ast }$, and $\overline{\mathcal{R}(T)}$ is the completion of $\mathcal{R}(T)$. In this case, a solution is given by 
\begin{equation}
\label{eq:Picard}
h=\sum_{i=1}^{\infty }\frac{1}{\sigma_{i}}\langle r_{0},\psi_{i}\rangle\varphi_{i}.  
\end{equation}
\end{theorem}

See \citep{kress2013linear} for a proof.

Throughout, we do not assume that the operator $T$ is injective, i.e., we do not assume that Equation \eqref{eq:IP1} has a unique solution. Rather, we focus our interest on the estimation of the solution $h_{0}$ with minimum norm, which we refer to as the $L^{2}$-minimal solution, i.e., satisfying 
\[
h_{0}=\argmin_{h\in L^{2}(P_{0,V_{h}}):Th=r_{0}}\Vert h\Vert_{2},
\]
where in the display above and throughout the rest of the exposition, $\Vert\cdot \Vert _{2}$ is the $L^{2}\left( P_{0}\right) $ norm.
For instance, in the NPIV model, this is tantamount to assuming that the underlying structural function is $L^{2}$-minimal. The minimum norm solution always exists and it is unique. It is equal to the residual from the $L^{2}(P_{0,V_h})$-projection of any solution $h$ of \eqref{eq:IP1} onto the null space $\mathcal{N}(T)$ of the operator $T.$ In fact, when $T$ is compact, the solution will be that in Equation \eqref{eq:Picard}.

A well-known challenge in solving integral equations of the form \eqref{eq:IP1} is that in many settings of interest, e.g., when $T$ is compact, the $L^{2}$-minimal solution $h_{0}$ is not a continuous function of $r_{0}$, which renders the problem \emph{ill-posed}. Specifically, it follows from Equation \eqref{eq:Picard} that a small deviation $\delta \psi _{i}$ from $r_{0}$ can be significantly amplified, as it corresponds to the solution $h_{0}+\left(\delta /\sigma _{i}\right) \varphi _{i}$, and the absolute value of $\delta/\sigma _{i}$ tends to infinity as $i$ increases because $\sigma _{i}$ tends to 0. This lack of continuity implies that consistent estimation of $h_{0}$ is not feasible without employing some regularization technique. Several different types of regularization techniques exist and are well-studied in the literature. These include spectral cutoff, Tikhonov, iterative Tikhonov and Landweber. We refer the readers to \citep{engl1996regularization} for a detailed discussion of regularization for inverse problems. In this work, we use iterative Tikhonov regularization, described in Section \ref{sec:DIVreg}.

When employing regularization techniques, regularity conditions are required for attaining a convergence rate on the estimation error. In standard non-parametric regression problems, regularity conditions in the form of assumptions on smoothness or complexity are imposed on the non-parametric regression function of interest. In contrast, in inverse problems, regularity conditions are required for both the $L^2$-minimal solution $h_{0}$, as well as the operator $T$. One often invoked regularity condition in the literature is the so-called $\beta$-source condition \citep{tautenhahn1996error, carrasco2007linear, darolles2011nonparametric, bennett2023source}, which relates the smoothness of the target parameter to the smoothing effect of the operator.

\begin{assumption}[$\beta $-source condition]
\label{assm:beta} 
There exists $w_{0}\in L^{2}(P_{0,V_{h}})$ with $\Vert w_{0}\Vert _{2}\leq B$, for some constant $B$, such that 
\[
h_{0}=(T^{\ast }T)^{\beta /2}w_{0}.
\]
In the case of compact operators, this can be written as 
\[
h_{0}\in \left\{ h=\sum_{i=1}^{\infty }\langle h,\varphi _{i}\rangle \varphi_{i}:\sum_{i=1}^{\infty }\frac{\langle h,\varphi _{i}\rangle ^{2}}{\sigma_{i}^{\beta }}\leq B^{2}\right\} ,
\]
where $(\sigma _{i},\varphi _{i})_{i=1}^{\infty }$ is the eigen system of $T^{\ast }T$.
\end{assumption}

For compact operators, the $\beta$-source condition requires that if the operator is highly smoothing, the target parameter must be sufficiently smooth to ensure that the summation $\sum_{i=1}^{\infty }\langle h_{0},\varphi _{i}\rangle ^{2}/\sigma _{i}^{\beta }$ is finite. The $\beta$-source condition can be leveraged to control the regularization bias as demonstrated in Theorems \ref{thm:mainNN1} and \ref{thm:mainNN} below.

\section{Debiased Ill-Posed Regression}
\label{sec:DIVreg}

For a function $h:\mathcal{V}_h\rightarrow\mathbb{R}$, define its (root mean squared) \emph{projected error} as 
\begin{align*}
{\E\left[\E\left[g_1(V)h(V_h)-g_0(V)\mid V_q\right]^2\right]}^{1/2}
&={\E\left[\E\left[g_1(V)h(V_h)-g_1(V)h_0(V_h)\mid V_q\right]^2\right]}^{1/2}\\
&=\|T(h-h_0)\|_2.
\end{align*}
Moreover, define the (root mean squared) \emph{source error} of $h$ as 
\[
{\E\left[\left\{h(V_h)-h_0(V_h)\right\}^2\right]}^{1/2}=\|h-h_0\|_2.
\]
Note that the conditional moment restriction in Equation \eqref{eq:cmrNN} (and \eqref{eq:IP1}) can be equivalently written as $\|T(h-h_0)\|_2=0$.

\begin{proposition}
\label{prop:Yh0}
The set of minimizers of $\|T(h-h_0)\|_2$ coincides with that of $\E[\{\E[g_1(V)h(V_h)\mid V_q]-g_0(V)\}^2]$.
\end{proposition}

Based on Proposition \ref{prop:Yh0}, we have
\[
h_0
\in\argmin_{h\in L^2(P_{0,V_{h}})}\psi(h),
\]
where, for any function $h\in L^2(P_{0,V_{h}})$, we define
\[
\psi(h):=\E\left[\left\{\E\left[g_1(V)h(V_h)\mid V_q\right]-g_0(V)\right\}^2\right]=
\E\left[\left\{(Th)(V_q)-g_0(V)\right\}^2\right]=\|Th-g_0\|_2^2.
\]
This observation as well as the minimality assumption suggests the following estimator for $h_0$:
\[
\hat h_\lambda^\text{B}=\argmin_{h\in\Hc}\E_n\left[\left\{(\hat Th)(V_q)-g_0(V)\right\}^2\right]+\lambda\E_n\left[h^2(V_h)\right],
\]
where $\hat T$ is an estimator of the operator $T$, $\E_n$ is the empirical mean operator, $\Hc\subset L^2(P_{0,V_{h}})$ is a user specified function class, and $\lambda$ is the regularization parameter that can depend on the sample size $n$. $\hat h_\lambda^\text{B}$ is referred to as the Tikhonov regularized estimator. 
The regularization term ensures converging towards the $L^2$-minimal solution, yet introduces regularization bias, which depends on the value of the hyper parameter $\lambda$. This Tikhonov regularized estimation is in fact the approach taken by \cite{hartford2017deep} and later by \cite{li2024regularized} to estimate the structural equation in the NPIV framework. Finite sample analysis of the estimator $\hat h_\lambda^\text{B}$ for the NPIV setting is provided in \citep{li2024regularized}.

We notice that in the approach above, first, the nuisance parameter $T$ is estimated with $\hat T$,  then $\hat T$ is used for estimating $\psi(h)$, which in turn is used for estimating $h_0$. Hence, a concern with this estimation approach is that it can be sensitive to the misspecification and/or slow convergence rate of the estimator $\hat T$ of $T$. Therefore, it is desired to reduce the sensitivity of the estimator of $\psi(h)$ to that of $T$. To this end, we propose to augment the objective function with a debiasing term. This can be achieved by leveraging the influence function of $\psi(h)$ which we derive below.

\begin{theorem}
\label{thm:IFNN}
In the nonparametric model, for any given function $h\in L^2(P_{0,V_{h}})$, the influence function of $\psi(h)$ is as follows.
\begin{align*}
\psi_{P_0}^1(V;h)
=&
\left\{\E[g_1(V)h(V_h)\mid V_q]-g_0(V)\right\}^2\\
&+2\left\{\E[g_1(V)h(V_h)\mid V_q]-\E[g_0(V)\mid V_q]\}\{g_1(V)h(V_h)-\E[g_1(V)h(V_h)\mid V_q]\right\}-\psi(h).
\end{align*}
\end{theorem}

Based on Theorem \ref{thm:IFNN}, for a given function $h\in L^2(P_{0,V_{h}})$, we propose the following estimator for $\psi(h)$:
\begin{align*}
\hat\psi(h)=\E_n\left[\left\{(\hat Th)(V_q)-g_0(V)\right\}^2+2\left\{(\hat Th)(V_q)-\hat r(V_q)\right\}\left\{g_1(V)h(V_h)-(\hat Th)(V_q)\right\}\right],
\end{align*}
where $\hat T\in\mathcal{T}$ and $\hat r\in\mathcal{R}$ are estimators for $T$ and $r_0$, respectively, which are obtained from an independent sample.

We have the following result regarding the structure of the bias of $\hat\psi(h)$.
\begin{proposition}[Bias structure]
\label{prop:biasNN}
For any given function $h\in L^2(P_{0,V_{h}})$, the bias of the estimator $\hat\psi(h)$ possesses the following structure.
\begin{align*}
\psi(h)-\E[\hat\psi(h)]
&=\E\left[\left\{(Th)(V_q)-(\hat Th)(V_q)\right\}^2
+2\left\{(Th)(V_q)-(\hat Th)(V_q)\right\}\left\{\hat r(V_q)-r_0(V_q)\right\}
\right]\\
&\le \|Th-\hat Th\|_2^2+2\|Th-\hat Th\|_2\|r-\hat r_0\|_2.
\end{align*}
\end{proposition}
Proposition \ref{prop:biasNN} demonstrates that the bias of the estimator $\hat\psi(h)$ is of second order. It also demonstrates that the estimator is robust to the misspecification of the nuisance function $r_0$, in the sense that it is (asymptotically) unbiased as long as the nuisance parameter $T$ is (asymptotically) correctly specified, even if the nuisance parameter $r_0$ is misspecified.

Equipped with the influence function-based estimator $\hat\psi(h)$, we propose to estimate the function $h_0$ using the following iterative Tikhanov regularized estimator.
\begin{equation}
\label{eq:DIVRNN}
\begin{aligned}
\hat h^{\text{IF}}_{\lambda,t}=\argmin_{h\in\Hc}\hat\psi(h)+\lambda\E_n\left[\{h(V_h)-\hat h^{\text{IF}}_{\lambda,t-1}\}^2\right].
\end{aligned}
\end{equation}
In our analyses in the following sections, we use $t=2$ and $\hat h^{\text{IF}}_{\lambda,0}=0$. Therefore, our estimator will be $\hat h^{\text{IF}}_{\lambda,2}$.
However, any initial function or number of iterations $t$ can be used. The drawback of using a large $t$ is that it leads to an exponentially large constant in our expression for convergence rate upper bound (see the proof of Theorem \ref{thm:mainNN}); hence we avoid such choices. The discussion of selecting the hyper-parameter $\lambda$ is postponed to Subsection \ref{sec:HPselect}.

We require the following realizability and boundedness conditions.
\begin{assumption}
\label{assm:bdd} 
$(i)$ $h_0$ belongs to $\Hc$.
$(ii)$ Function classes $\Hc$ and $\mathcal{R}$ are uniformly bounded with respect to $L^2(P_0)$ norm.
\end{assumption}

\subsection{Convergence Rate}
\label{sec:mainres}

In this subsection, we present our results regarding the convergence rate of the projected error and the source error of the estimator $\hat h^{\text{IF}}_{\lambda,2}$. 
We use the statistical learning-theoretic local complexity measures of localized Rademacher complexity and the related concept of critical radius for characterizing the complexity of the function space and convergence rate of excess risk \citep{wainwright2019high}. Before presenting our result, we briefly review these notions.

\begin{definition}[Localized Rademacher complexity]
For a given $\delta>0$, and function class $\mathcal{H}$,  the localized Rademacher complexity of $\mathcal{H}$ is defined as
\[
\mathcal{R}(\delta,\mathcal{H})\coloneqq
\E_{\epsilon,V_h}\Bigg[ \sup_{\overset{h\in\mathcal{H}}{\|h\|_{2}\le \delta}} \Big|  \frac{1}{n}\sum_{i=1}^n\epsilon_i h(V_{h,i}) \Big| \Bigg],
\]
where $\{V_{h,i}\}_{i=1}^n$ are i.i.d. samples from the underlying distribution and $\{\epsilon_i\}_{i=1}^n$ are i.i.d. Rademacher variables taking values in $\{-1,+1\}$ with equal probability, independent of $\{V_{h,i}\}_{i=1}^n$.
\end{definition}
\begin{definition}[Critical radius]
The critical radius of a function class $\mathcal{H}$, denoted by $\delta_n^*$, is the smallest solution to the inequality $\mathcal{R}(\delta,\mathcal{H})\le\delta^2$.
\end{definition}
\cite{wainwright2019high} also provided the empirical counterparts of the localized Rademacher complexity and critical radius, which can be used to estimate the corresponding true values.

In the sequel, the notation $X_1\lesssim X_2$ means that there exists a constant $c$ such that $X_1\le c X_2$ ($X_1\gtrsim X_2$ is defined similarly), and the notation $X_1\asymp X_2$ means that there exist constants $c_1$ and $c_2$ such that $c_1X_1\le X_2\le c_2X_1$. Moreover, for a function class $\mathcal{F}$, we define $\mathcal{F}-\mathcal{F}:=\{f_1-f_2:f_1,f_2\in\mathcal{F}\}$, and denote the star hull of $\mathcal{F}$ by $star(\mathcal{F})$ which is defined as $\{\alpha f: f\in \mathcal{F}, \alpha\in[0,1]\}$.

We are now ready to present our results regarding the convergence rate of the projected and source errors of estimator $\hat h^{\text{IF}}_{\lambda,2}$. Our first result provides a bound on the projected error. We will use the following function classes in this result: $\mathcal{G}_1:=\{\hat Th:h\in\mathcal{H}\}$, and $\mathcal{G}_2:=\{\tilde Th_0:\tilde T\in\mathcal{T}\}$. Our first result does not utilize the $\beta$-source condition and only focuses on projected error.

\begin{theorem}
\label{thm:mainNN0}
\begin{sloppypar}
Suppose Assumptions \ref{assm:exists} and \ref{assm:bdd} hold. 
Let $\delta_{n}$ be an upper bound on the critical radius of $star(\Hc-\Hc)$, and 
$\delta_{M,n}$ be an upper bound on the critical radii of $star(\mathcal{G}_1-\mathcal{G}_1)$, $star(\mathcal{G}_2-\mathcal{G}_2)$, and $star(\mathcal{R}-\mathcal{R})$, which satisfy $\delta_{n}^2,\delta_{M,n}^2\gtrsim  \frac{\log(\log(n))+\log(1/\zeta)}{n}$.
\end{sloppypar}
Define
\[
\Delta_{M,n}=\max\left\{\delta_{M,n}^2,\|T-\hat T\|^2,\|\hat r-r_0\|_2^2\right\},
\]
Then, 
for any $0<\lambda$, with probability at least $1-5\zeta$, for $n$ with $\delta_n<1$, we have
\begin{align*}
\|T(\hat h^{\text{IF}}_{\lambda,2}-h_0)\|_2^2
\lesssim
\Delta_{M,n}
+\lambda\delta_n^2+
\lambda\|\hat h^{\text{IF}}_{\lambda,2}-h_0\|_2.
\end{align*}
In particular, with a value $\lambda^*\lesssim\Delta_{M,n}$, with probability at least $1-5\zeta$, we have
\begin{align*}
\|T(\hat h^{\text{IF}}_{\lambda^*,2}-h_0)\|_2^2
\lesssim
\Delta_{M,n}.
\end{align*}
\end{theorem}

Next, we show that by additionally requiring Assumption \ref{assm:beta} (the $\beta$-source condition), we can obtain an alternative bounds on the projected error, and we can also bound the source error. We will use the function $h_{\lambda,t}^*$ defined below in our result.
\begin{equation}
\label{eq:popobjNN}
\begin{aligned}
h^*_{\lambda,t}=\argmin_h\E\left[\left\{(Th)(V_q)-g_0(V)\right\}^2\right]+\lambda\E\left[\{h(V_h)-h^*_{\lambda,t-1}(V_h)\}^2\right],
\end{aligned}
\end{equation}
with $h^*_{\lambda,0}=0$.
We present two versions of this result in Theorems \ref{thm:mainNN1} and \ref{thm:mainNN} below. While one can simply take the minimum of the bounds in these theorems, we state them separately to maintain clarity in the exposition.

\begin{theorem}
\label{thm:mainNN}
Suppose Assumptions \ref{assm:exists}, \ref{assm:beta}, and \ref{assm:bdd} hold, and the functions $h_{\lambda,1}^*$ and $h_{\lambda,2}^*$ defined in \eqref{eq:popobjNN} belong to $\mathcal{H}$. 
Let $\delta_n$ be an upper bound on the critical radius of $star(\Hc-\Hc)$, which satisfies $\delta_n^2\gtrsim  \frac{\log(\log(n))+\log(1/\zeta)}{n}$.
Define
\[
\Delta_n=\max\left\{\delta_n^2,\|T-\hat T\|^4,\|T-\hat T\|^2\|\hat r-r_0\|_2^2\right\},
\]
\begin{sloppypar}
\end{sloppypar}
Then, 
for any $0<\lambda$ with probability at least $1-2\zeta$, we have
\begin{align*}
&\|T(\hat h^{\text{IF}}_{\lambda,2}-h_0)\|_2^2
\lesssim
\frac{1}{\lambda}
\Delta_n
+
\lambda^{\min\{4,\beta+1\}},\\
&\|\hat h^{\text{IF}}_{\lambda,2}-h_0\|_2^2
\lesssim
\frac{1}{\lambda^2}
\Delta_n
+
\lambda^{\min\{4,\beta\}}.
\end{align*}
In particular, with a value $\lambda^*\asymp\Delta_n^{\frac{1}{\min\{5,\beta+2\}}}$, with probability at least $1-2\zeta$, for $n$ with $\Delta_n<1$, we have
\begin{align*}
&\|T(\hat h^{\text{IF}}_{\lambda^*,2}-h_0)\|_2^2
\lesssim
\Delta_n^{\frac{\min\{4,\beta+1\}}{\min\{5,\beta+2\}}},\\
&\|\hat h^{\text{IF}}_{\lambda^*,2}-h_0\|_2^2
\lesssim
\Delta_n^{\frac{\min\{3,\beta\}}{\min\{5,\beta+2\}}}.
\end{align*}
\end{theorem}

\begin{theorem}
\label{thm:mainNN1}
Suppose Assumptions \ref{assm:exists}, \ref{assm:beta}, and \ref{assm:bdd} hold, and the functions $h_{\lambda,1}^*$ and $h_{\lambda,2}^*$ defined in \eqref{eq:popobjNN} belong to $\mathcal{H}$.
Let $\delta_{n}$ be an upper bound on the critical radius of $star(\Hc-\Hc)$, and 
$\delta_{M,n}$ be an upper bound on the critical radii of $star(\mathcal{G}_1-\mathcal{G}_1)$, $star(\mathcal{G}_2-\mathcal{G}_2)$, and $star(\mathcal{R}-\mathcal{R})$, which satisfy $\delta_{n}^2,\delta_{M,n}^2\gtrsim  \frac{\log(\log(n))+\log(1/\zeta)}{n}$.
Define
\[
\Delta_{M,n}=\max\left\{\delta_{M,n}^2,\|T-\hat T\|^2,\|\hat r-r_0\|_2^2\right\},
\]
Then, for any $0<\lambda<1$, with probability at least $1-5\zeta$, we have
\begin{align*}
&\|T(\hat h^{\text{IF}}_{\lambda,2}-h_0)\|_2^2
\lesssim
\Delta_{M,n}
+
\lambda\delta_n^2+\lambda^{\min\{2,\beta+1\}},\\
&\|\hat h^{\text{IF}}_{\lambda,2}-h_0\|_2^2
\lesssim
\frac{1}{\lambda}\Delta_{M,n}
+
\delta_n^2+\lambda^{\min\{1,\beta\}}.
\end{align*}
In particular, with a value $\lambda^*\asymp\Delta_{M,n}^{\frac{1}{\min\{2,\beta+1\}}}$, with probability at least $1-5\zeta$, we have
\begin{align*}
&\|T(\hat h^{\text{IF}}_{\lambda^*,2}-h_0)\|_2^2
\lesssim
\Delta_{M,n}+\delta_n^2,\\
&\|\hat h^{\text{IF}}_{\lambda^*,2}-h_0\|_2^2
\lesssim
\Delta_{M,n}^{\frac{\min\{1,\beta\}}{\min\{2,\beta+1\}}}+\delta_n^2.
\end{align*}
\end{theorem}

Several remarks are in order.

\begin{remark}
It is important to note the difference between estimators $\hat h^{\text{IF}}_{\lambda,2}$ and $\hat h^{\text{B}}_{\lambda,2}$. The estimator $\hat h^{\text{IF}}_{\lambda,2}$ requires the estimation of one extra nuisance function, namely, $\hat r$. This nuisance function is leveraged for debiasing the baseline estimator of $\psi(h)$. \cite{li2024regularized} analyzed the convergence rate of $\hat h^{\text{B}}_{\lambda,2}$ (for the NPIV setting), which does not contain the debiasing step. The authors showed that, under the $\beta$-source condition, with high probability, $\hat h^{\text{B}}_{\lambda,2}$ satisfies (in our notation)
\begin{align*}
\|\hat h_{\lambda,2}^\text{B}-h_0\|_2^2
\lesssim
\left(	
\frac{1}{\lambda^2}
\max
\left\{\delta_n^2,\|T-\hat T\|^2\right\}
+
\lambda^{\min\{4,\beta\}}\right).
\end{align*}
We note that $T$ is of the form of a conditional expectation operator and is expected to be a challenging nuisance component to be estimated. Hence, the term $\|T-\hat T\|^2$ is likely to dominate $\delta_n^2$. Compared to this rate, the convergence rate in our result in Theorem \ref{thm:mainNN} depends on $\|T-\hat T\|^4$ and $\|T-\hat T\|^2\|\hat r-r_0\|_2^2$. Hence, it can significantly improve the convergence rate as it resolves the concern regarding a slow convergence rate of $\hat T$.
\end{remark}

\begin{remark}
\label{rmk:robust}
As mentioned earlier, we require the estimation of one extra nuisance function, namely, $r_0$. Importantly, our estimator under $\beta$-source condition is indeed robust to the misspecification of this extra nuisance function. Formally, if we misspecify $r_0$, for the rate in Theorem \ref{thm:mainNN} we will have
\[
\max\left\{\delta_n^2,\|T-\hat T\|^4,\|T-\hat T\|^2\|\hat r-r_0\|_2^2\right\}
\lesssim\max\left\{\delta_n^2,\|T-\hat T\|^2\right\}.
\]
\end{remark}
Therefore, in the worst case, we are back to the result of  \cite{li2024regularized}.

\begin{remark}
We note the presence of the term $\lambda^{\min\{4,\beta\}}$ in our result in Theorem \ref{thm:mainNN}, which is inherited from the regularization bias in the convergence rate. This term shows that if the ill-posedness is mild such that $\beta>4$, we will not benefit from this mild ill-posedness. This effect is related to the value of the so-called ``qualification'' of the regularization method. 
Importantly, this term is an already improved version of $\lambda^{\min\{2,\beta\}}$ which would have appeared in the result if we used basic Tikhonov regularization instead of our 2-iteration Tikhonov regularization, and it can be further improved and replaced with $\lambda^{\beta}$ if we increase the number of iterations in the iterative Tikhonov regularization. However, as mentioned earlier, $t$-iteration Tikhonov regularization will lead to a constant exponential in $t$ in the upper bound in the existing analysis methods in the literature. Therefore, we chose to only use 2 iterations. As will be seen in Section \ref{sec:linfuncal}, this choice suffices for obtaining root-$n$ consistent estimators for linear functionals of the ill-posed regression functions.
See \citep{engl1996regularization, cavalier2011inverse} for further detailed discussions, derivations, and analyses of iterative Tikhonov regularization. Also see \citep{bennett2023source,li2024regularized} for the form of the exponential constant that can arise with $t$-iteration Tikhonov regularization.
\end{remark}

\begin{remark}
In a related work, \cite{foster2019orthogonal} proposed a general framework for statistical learning in the presence of nuisance functions. They demonstrate that if the population loss function satisfies a condition called Neyman orthogonality, the impact of the nuisance estimation error on the excess risk bound is of second order. Compared to that work, we use a loss function derived from the influence function of an initial loss, which has more structure: by virtue of using the influence function, our loss function also satisfied Neyman orthogonality, yet at any function $h$, not just the minimal solution $h_0$. Moreover, our loss function has the desirable robustness property, stated in Remark \ref{rmk:robust}, which a Neyman orthogonal loss function does not necessarily possess. In addition, we note that the model with conditional moment restriction \eqref{eq:cmrNN} does not directly restrict the function $h$, and hence compared to the setting in \citep{foster2019orthogonal}, we deal with the additional challenge of solving an ill-posed inverse problem. Finally, that work does not consider hyper-parameter tuning in their framework, which is one of our main focuses in this work, studied in the following subsection.
\end{remark}

\subsection{Hyper-Parameter Selection}
\label{sec:HPselect}

Although Theorems \ref{thm:mainNN} and \ref{thm:mainNN1} provide  bounds on the source error, the value of the hyper parameter $\lambda^*$ in those theorems depends on the parameter $\beta$, an upper bound on the critical radii of certain function classes, and convergence rates of $\hat T$ and $\hat r$. Most of these quantities are usually unknown.
In this subsection, we address hyper-parameter selection for the debiased ill-posed regression and demonstrate the resulting potential rate loss due to not knowing $\lambda^*$. Our results can be considered as generalization of existing results in the literature on rate analysis for hyper-parameter tuning (e.g., \citep{vaart2006oracle, van2003unified}) to the setting of ill-posed regression.

Let $\mathcal{C}:=\{\hat h_1,...,\hat h_M\}$ be our candidate set containing $M$ candidate estimators for the function $h_0$. For any given function $h$, recall the definition
\begin{align*}
\hat\psi(h)
&=\E_n\left[\left\{(\hat Th)(V_q)-g_0(V)\right\}^2+2\left\{(\hat Th)(V_q)-\hat r(V_q)\right\}\left\{g_1(V)h(V_h)-(\hat Th)(V_q)\right\}\right].
\end{align*}
Suppose candidates in $\mathcal{C}$ and nuisance components $\hat T$ and $\hat r$ are learned on separate independent data folds. We propose to select an estimator candidate $\hat h$ on an independent validation dataset (the empirical expectation above is with respect to the validation data) as follows:
\begin{equation}
\label{eq:mainest}
\hat h:=\argmin_{h\in\mathcal{C}}\hat\psi(h).
\end{equation}
Compared to the estimator in \eqref{eq:DIVRNN}, here we do not have the regularization term and the optimization is over $\mathcal{C}$ as opposed to $\mathcal{H}$.

Our goal is to analyze the convergence rate of the source error and the projected error of $\hat h$ in \eqref{eq:mainest}. We start with the analysis of the projected error of $\hat h$.

\begin{theorem}
\label{thm:hpc}
Suppose Assumptions \ref{assm:exists} and \ref{assm:bdd} hold.
Let $h^*\in\argmin_{h\in\mathcal{C}}\|T(h-h_0)\|_2^2$.
Then, for the projected error, with probability at least $1-3\zeta$, our pick $\hat h$ satisfies
\begin{align*}
\big\|T(\hat h-h_0)\big\|_2^2
&\lesssim
\big\|T(h^*-h_0)\big\|_2^2\\
&\quad+\frac{\log (2M/\zeta)}{n}+\max_{h\in\mathcal{C}}\| h-h^*\|_2\|T-\hat T\|^2
+\max_{h\in\mathcal{C}}\| h-h^*\|_2^2\|\hat r-r_0\|_2^2.
\end{align*}
\end{theorem}

In the context of hyper-parameter selection for the estimator in \eqref{eq:DIVRNN}, suppose we have $M$ candidates $\lambda_1<\lambda_2<\cdots<\lambda_M$ for the hyper-parameter in the optimization problem \eqref{eq:DIVRNN}. Let $\hat h_i$ denote the estimator corresponding to $\lambda_i$, for $i\in\{1,...,M\}$; let $\mathcal{C}:=\{\hat h_1,...,\hat h_M\}$ be the candidate set.
We again consider candidates in $\mathcal{C}$ and nuisance components $\hat T$ and $\hat r$ are learned on separate independent data folds, and we select an estimator candidate on an independent validation dataset based on the following approach. 
Let $(b_n)_{n=1}^\infty$ and $(B_n)_{n=1}^\infty$ be positive decreasing sequences converging to zero, approximating $\delta_n^{1-\varepsilon}$ and $\delta_n^{1/3}$, respectively, where $\delta_n$ is defined in Theorem \ref{thm:mainNN}, and $0<\varepsilon\ll1$. 
We choose $M=n$, and
\[
\lambda_i=b_n+\frac{i}{n}B_n ,\qquad i\in\{1,...,n\},
\]
where $n$ is the sample size for both training and validation datasets.
We assume that the debiasing succeed in the sense that $\delta^2_n=\Delta_n$ and $\delta^2_{M,n}=\Delta_{M,n}$, where $\delta_n$, $\delta_{M,n}$, $\Delta_n$, and $\Delta_{M,n}$ were defined in the statement of Theorems \ref{thm:mainNN} and \ref{thm:mainNN1}. We also consider $\Delta_{M,n}\lesssim\delta_n$. In this case, our choices for $(b_n)_{n=1}^\infty$ and $(B_n)_{n=1}^\infty$ guarantee that our upper bounds in Theorems \ref{thm:mainNN} and \ref{thm:mainNN1} converge to zero. We only utilize knowledge of $\delta_n$. However, if one has knowledge of other components involved in $\Delta_n$ and $\Delta_{M,n}$, it can be incorporated to relax the aforementioned assumption.

We have the following corollary of Theorems \ref{thm:mainNN}-\ref{thm:hpc}.

\begin{corollary}
\label{cor:thm}
Let $\lambda^\dagger:=\argmin_{\lambda\in\{\lambda_1,...,\lambda_n\}}\|T(\hat h_{\lambda,2}^\text{IF}-h_0)\|_2^2$, and let $h^*:=\hat h_{\lambda^\dagger,2}^\text{IF}$.
Suppose Assumptions \ref{assm:exists}-\ref{assm:bdd} hold.  
Then, for the projected error, with probability at least $1-5\zeta$, our pick $\hat h$ satisfies
\begin{align*}
\big\|T(\hat h-h_0)\big\|_2^2
&\lesssim
\min\Big\{\Delta_n^{\frac{\min\{4,\beta+1\}}{\min\{5,\beta+2\}}},
\Delta_{M,n}+\delta_n^2
\Big\}\\
&\quad
+\frac{\log (2n/\zeta)}{n}+\max_{h\in\mathcal{C}}\| h-h^*\|_2\|T-\hat T\|^2
+\max_{h\in\mathcal{C}}\| h-h^*\|_2^2\|\hat r-r_0\|_2^2.
\end{align*}
\end{corollary}

Several remarks are in order.
\begin{remark}
Comparing the rate corresponding to $\lambda^*$ for the projected error in Theorems \ref{thm:mainNN} and \ref{thm:mainNN1} with the rate in Corollary \ref{cor:thm}, we see that when the first term dominates in the latter, our hyper-parameter tuning does not degrade the convergence rate. However, if either of the last two terms dominates, a potential loss in rate may occur. Whether this loss can be mitigated through alternative bounding techniques or tuning strategies remains an open question for future research. 
\end{remark}

\begin{remark}
As mentioned earlier, our choices for $(b_n)_{n=1}^\infty$ and $(B_n)_{n=1}^\infty$ guarantee that our upper bounds in Theorems \ref{thm:mainNN} and \ref{thm:mainNN1} converge to zero. That is, $\max_{h\in\mathcal{C}}\| h-h^*\|_2^2$ will converge to zero, and hence, even if $\hat r$ is misspecified, the term $\max_{h\in\mathcal{C}}\| h-h^*\|_2^2\|\hat r-r_0\|_2^2$ vanishes asymptotically and hence, misspecification of $\hat r$ does not affect consistency.
\end{remark}

\begin{remark}
Suppose the operator $T$ were known. In this case, the statistical model, and hence, the influence function derived in Theorem \ref{thm:IFNN} will be different. Specifically, the influence function will not have the debiasing term. In this case, the counterpart of Corollary \ref{cor:thm} will not have the last two terms. However, if one mistakenly uses the estimator containing the debiasing term, then they will have the extra term $\max_{h\in\mathcal{C}}\| h-h^*\|_2^2\|\hat r-r_0\|_2^2$. 
As mentioned in the previous remark, this will not affect consistency. However, it may result in a weaker bound on the projected risk compared to an estimator that does not employ the debiasing term. It is not clear whether this can be mitigated through alternative bounding techniques or tuning strategies. We leave investigating the answer to this question to future work.
\end{remark}

For analyzing the source error, we require the following additional assumption.

\begin{assumption}[$\alpha$-error condition]
\label{assm:alpha}
There exists constants $0<\alpha<\infty$ and $0<C_\alpha<\infty$, a sequence $0<(\mu_i^{(n)})_{i,n=1}^\infty<1$, indexed by $i$ and $n$, which is non-increasing in $i$ for all $n$ and non-increasing in $n$ for all $i$, and a basis for $L^2(P_{0,V_h})$, $(\phi_i)_{i=1}^\infty$, such that for all $h_n\in\mathcal{C}$ (obtained from sample of size $n$), with error function defined as $e_n:=h_n-h_0$, we have
\begin{enumerate}
	\item the error function is smooth in the sense that
	\[
	\sum_{i=1}^\infty\frac{\langle e_n,\phi_i\rangle^2}{\mu_i^{(n)}}\le C_\alpha,
	\]
	\item the projected error of $h_n$ can be lower bounded as
	\[
	\sum_{i=1}^\infty (\mu_i^{(n)})^\alpha \langle e_n,\phi_i\rangle^2\le \|Te_n\|^2_2.
	\]
\end{enumerate}
\end{assumption}

Assumption \ref{assm:alpha} is reminiscent of those used in analyses based on Hilbert scales (e.g., \citep{chen2011rate, florens2011identification}).
Roughly speaking, Assumption \ref{assm:alpha} aims to characterize the smoothness and the convergence (if the projected error converges) of the error functions with one constant $\alpha$, hence the name $\alpha$-error condition. The assumption requires a certain level of \emph{quality} on the candidates in $\mathcal{C}$ (not on all functions in the function class $\mathcal{H}$) and is essentially that the smoothness and convergence is so that the \emph{same} value of $\alpha$ works for all sample sizes. 
In general, one can find a value $\alpha$ only if the sequence $(\mu_i^{(n)})_{i,n=1}^\infty$ has an overall decreasing trend and converges to zero (see Remark \ref{rmk:abcond} for a special case where $(\mu_i^{(n)})_{i,n=1}^\infty$ need not converge to zero in $n$). 
In such general cases, if convergence of error function is too slow (captured by $\mu_i^{(n)}\rightarrow 0$ too slowly as $n\rightarrow\infty$), or if the smoothness is not sufficient (captured by $\mu_i^{(n)}\rightarrow 0$ too slowly as $i\rightarrow\infty$), then we will not be able to find a finite constant $\alpha$.

\begin{remark}
As mentioned earlier, in general, one can find a value $\alpha$ only if the sequence $(\mu_i^{(n)})_{i,n=1}^\infty$ converges to zero as $n$ increases, which in turn implies that Assumption \ref{assm:alpha} is well-justified if the source error of the candidates converge to zero. We saw in Theorems \ref{thm:mainNN} and \ref{thm:mainNN1} that $\beta$-source condition can render this requirement feasible. 
\end{remark}

\begin{remark}
\label{rmk:abcond}
An important instantiation of Assumption \ref{assm:alpha} is in the case of injective operators. 
Consider the case that the operator $T$ is injective and compact, with singular system $(\sigma _{i},\varphi _{i},\psi _{i})_{i=1}^{\infty }$. In this case, $(\varphi _{i})_{i=1}^{\infty }$ is a basis for $L^2(P_{0,V_h})$. Define $(\tilde\sigma _{i})_{i=1}^{\infty }$ as $(\sigma _{i})_{i=1}^{\infty }$ normalized by the largest singular value. Then, a special version of Assumption \ref{assm:alpha} is to assume 
	\[
	\sum_{i=1}^\infty\frac{\langle e_n,\varphi_i\rangle^2}{\tilde\sigma_i^\beta}\le C_\alpha,
	\]
	which is a modified version of the $\beta$-source condition, but for error functions, as opposed to for $h_0$. Note that this will also lead to 
	\[
	\sum_{i=1}^\infty \tilde\sigma_i^{\alpha\beta} \langle e_n,\varphi_i\rangle^2\lesssim \|Te_n\|^2_2=\sum_{i=1}^\infty \sigma_i^2 \langle e_n,\varphi_i\rangle^2,
	\]
as long as $\alpha\ge2/\beta$.
Therefore, the sequence $0<(\mu_i^{(n)})_{i,n=1}^\infty<1$ in this case is constant in $n$.
Note that in this case, the assumption essentially boils down to requiring convergence of the errors to zero and requiring that the sample size is large enough so that the errors are smooth.
\end{remark}

\begin{example}
	Suppose $\mu_i^{(n)}=1/(n^{0.1}i^3)$, and for all candidates, $\|Te_n\|^2_2\gtrsim n^{-1/3}$. Then
	\begin{align*}
		\sum_{i=1}^\infty (\mu_i^{(n)})^\alpha \langle e_n,\phi_i\rangle^2
		&\le n^{\frac{-\alpha}{10}}\left(\sum_{i=1}^\infty\frac{1}{i^{6\alpha}}\right)^{\frac{1}{2}}\left(\sum_{i=1}^\infty\langle e_n,\phi_i\rangle^4\right)^{\frac{1}{2}}\\
		&\le n^{\frac{-\alpha}{10}}\left(\frac{\pi^3}{\sqrt{945}}\right)\|e_n\|_2.
	\end{align*}
	Therefore, it suffices that we have
	\[
	n^{\frac{-\alpha}{10}}\|e_n\|_2\lesssim n^{-1/3},
	\]
	or equivalently
	\[
	\|e_n\|_2\lesssim n^{\frac{\alpha}{10}-\frac{1}{3}},
	\]
	which is achieved by $\alpha=10/3$ as long as the source error of the candidates are converging to zero.
\end{example}

In the following, we present our result regarding the convergence rate of the source error of the estimator $\hat h$ in \eqref{eq:mainest}.

\begin{theorem}
\label{thm:hpcs}
Under Assumption \ref{assm:alpha} the estimator $\hat h$ satisfies
\[
\|\hat h-h_0\|_2^{2}\lesssim\|T(\hat h-h_0)\|_2^{\frac{2}{1+\alpha}}.
\]
\end{theorem}

\section{Linear Functionals of Ill-Posed Regression Functions}
\label{sec:linfuncal}

In a recent work, \cite{ghassami2022minimax} considered a parameter class of regular parameters $\psi_0$ with influence function of the form 
\[
IF_{\psi_0}(V)=s_1(V)q_0(V_q)h_0(V_h)+s_2(V)q_0(V_q)+s_3(V)h_0(V_h)+s_4(V)-\psi_0,
\]
where $s_1$, $s_2$, $s_3$, and $s_4$ are known functions. 
This is a generalization of the Robins' parameter class \citep{robins2008higher} to the case where $V_h\neq V_q$.
\cite{ghassami2022minimax} showed that the nuisance functions $h_0$ and $q_0$ are solutions to conditional moment restrictions of the form \eqref{eq:cmrNN}. Specifically, $h_0$ solves
\[
\E\left[s_1(V)h_0(V_h)+s_2(V)\mid V_q\right]=0,
\]
and $q_0$ solves
\[
\E\left[s_1(V)q_0(V_q)+s_3(V)\mid V_h\right]=0.
\]
The authors used an adversarial learning approach to design estimators $\hat h$ and $\hat q$ for $h_0$ and $q_0$, respectively. These estimators are in-turn used in a debiased (influence function-based) estimator of $\psi_0$ of the form
\begin{equation}
\label{eq:estclass}
\hat\psi=\E_n\left[
s_1(V)\hat q(V_q)\hat h(V_h)+s_2(V)\hat q(V_q)+s_3(V)\hat h(V_h)+s_4(V)
\right].
\end{equation}
Here, we apply our ill-posed regression strategy to estimate $h_0$ and $q_0$, and extend the debiasing point of view by also using a debiased (influence function-based) estimator for the nuisance functions. Hence, we propose an estimation strategy for $\psi_0$ with two layers of debiasing.

It can be shown that the estimator $\hat\psi$ has the so-called mixed-bias property \citep{rotnitzky2021characterization}, which implies that the bias in estimating $\psi_0$ is connected to a product of the biases in estimating $h_0$ and $q_0$. Formally, the mixed-bias property states that the bias of the estimator $\hat\psi$ satisfies
\[
\psi_0-\E[\hat\psi]
=\E[s_1(V)(\hat q(V_q)-q_0(V_q))(\hat h(V_h)-h_0(V_h))].
\]
This mixed-bias property results in the estimator to be doubly robust, meaning that it is (asymptotically) unbiased as long as either $\hat h$ or $\hat q$ is (asymptotically) correctly specified.
\cite{ghassami2022minimax} used the mixed-bias property to study the convergence rate of $\hat\psi$ and showed that, under certain mild regularity conditions and consistency of the estimators $\hat h$ and $\hat q$, the estimator $\hat\psi$ is $\sqrt{n}$-consistent and asymptotically normal if 
\begin{equation}
\label{eq:rootn}
\begin{aligned}
\min\Big\{&\big\|\hat q-q_0\big\|_2\big\|\E[s_1(V)\hat h(V_h)-s_1(V)h_0(V_h)\mid V_q]\big\|_2,\\
&~~~\big\|\hat h-h_0\big\|_2\big\|\E[s_1(V)\hat q(V_q)-s_1(V)q_0(V_q)\mid V_h]\big\|_2\Big\}=o_p(n^{-1/2}).
\end{aligned}
\end{equation}
That is, the rate requirement is on the product of the projected error of one of the nuisance functions and the source error of the other nuisance function.

Let $T_h:L^2(P_{0,V_h})\rightarrow L^2(P_{0,V_q})$ be the linear, bounded, operator $(T_hh)(V_q)=\E[s_1(V)h(V_h)\mid V_q]$, and let $r_h\in L^2(P_{0,V_q})$ be defined as $r_h(V_q)=\E[s_2(V)\mid V_q]$.
Moreover, 
let $T_q:L^2(P_{0,V_q})\rightarrow L^2(P_{0,V_h})$ be the linear, bounded, operator $(T_qq)(V_h)=\E[s_1(V)q(V_q)\mid V_h]$, and let $r_q\in L^2(P_{0,V_h})$ be defined as $r_q(V_h)=\E[s_3(V)\mid V_h]$.
We have the following corollary regarding the root-$n$ consistency of the estimator $\hat\psi$.

\begin{corollary}
\label{cor:1m}
Suppose Assumptions \ref{assm:exists}-\ref{assm:bdd} hold with the smoothness parameter $\beta_h$ (resp. $\beta_q$). Let $\hat h_{{\lambda_h^*,2}}^{IF}$ (resp. $\hat q_{{\lambda_q^*,2}}^{IF}$) be the proposed estimator based on the debiased ill-posed regression strategy for the nuisance function $h_0$ (resp. $q_0$), with function class $\Hc$ (resp. $\mathcal{Q}$) with an upper bound $\delta_{\mathcal{H},n}$ (resp. $\delta_{\mathcal{Q},n}$) on its critical radius as described in Theorem \ref{thm:mainNN}. 
Define 
\begin{align*}
&\Delta_{h,n}:=
\max\{\|T_h-\hat T_h\|^4, \|T_h-\hat T_h\|^2\|\hat r_h-r_h\|_2^2,\delta_{\mathcal{H},n}^2\},\\
&\Delta_{q,n}:=
\max\{\|T_q-\hat T_q\|^4, \|T_q-\hat T_q\|^2\|\hat r_q-r_q\|_2^2, \delta_{\mathcal{Q},n}^2\}.
\end{align*}
With 
$
\lambda^*_h\asymp\Delta_{h,n}^{\frac{1}{\min\{5,\beta_h+2\}}}
\text{ and }
\lambda^*_q\asymp\Delta_{q,n}^{\frac{1}{\min\{5,\beta_q+2\}}},
$
Condition \eqref{eq:rootn} is satisfied with high probability if
\begin{equation}
\label{eq:corcond}
\begin{aligned}
\min\Bigg\{
\Delta_{q,n}^{\frac{\min\{3,\beta_q\}}{2\min\{5,\beta_q+2\}}}
\Delta_{h,n}^{\frac{\min\{4,\beta_h+1\}}{2\min\{5,\beta_h+2\}}},
\Delta_{h,n}^{\frac{\min\{3,\beta_h\}}{2\min\{5,\beta_h+2\}}}
\Delta_{q,n}^{\frac{\min\{4,\beta_q+1\}}{2\min\{5,\beta_q+2\}}}
\Bigg\}
=o_p(n^{-1/2}).
\end{aligned}
\end{equation}
\end{corollary}
\begin{corollary}
\label{cor:2m}
Consider the assumptions and notations in Corollary \ref{cor:1m}. Suppose in addition, Assumption \ref{assm:alpha} holds with parameter $\alpha_h$ (resp. $\alpha_q$). Let $\hat h$ (resp. $\hat q$) be the proposed cross-validated estimator for the nuisance function $h_0$ (resp. $q_0$) chosen from candidate set $\mathcal{C}_h$ (resp. $\mathcal{C}_q$).
Define 
\begin{align*}
&\Theta_{h,n}:=
\frac{\log (2n/\zeta)}{n}
+\|T_h-\hat T_h\|^2
+\|\hat r_h-r_h\|_2^2
+\Delta_{\mathcal{H},n}^{\frac{\min\{4,\beta_h+1\}}{\min\{5,\beta_h+2\}}},\\
&\Theta_{q,n}:=
\frac{\log (2n/\zeta)}{n}
+\|T_q-\hat T_q\|^2
+\|\hat r_q-r_q\|_2^2
+\Delta_{\mathcal{Q},n}^{\frac{\min\{4,\beta_q+1\}}{\min\{5,\beta_q+2\}}}.
\end{align*}
Condition \eqref{eq:rootn} is satisfied with high probability if
\begin{equation}
\label{eq:corcond2}
\begin{aligned}
\min\Bigg\{
\Theta_{q,n}^{\frac{1}{2+2\alpha_q}}
\Theta_{h,n}^{\frac{1}{2}},
\Theta_{h,n}^{\frac{1}{2+2\alpha_h}}
\Theta_{q,n}^{\frac{1}{2}}
\Bigg\}
=o_p(n^{-1/2}).
\end{aligned}
\end{equation}
\end{corollary}

\begin{example}
Suppose 
\begin{align*}
\rho_n&\asymp\|T_h-\hat T_h\|^{\frac{\min\{5,\beta_h+2\}}{\min\{4,\beta_h+1\}}}
\asymp\|\hat r_h-r_h\|_2^{\frac{\min\{5,\beta_h+2\}}{\min\{4,\beta_h+1\}}}
\asymp\delta_{\mathcal{H},n}\\
&\asymp\|T_q-\hat T_q\|^{\frac{\min\{5,\beta_h+2\}}{\min\{4,\beta_h+1\}}}
\asymp\|\hat r_q-r_q\|_2^{\frac{\min\{5,\beta_h+2\}}{\min\{4,\beta_h+1\}}}
\asymp\delta_{\mathcal{Q},n}.
\end{align*} 
Furthermore, suppose $\beta_h=\beta_q=3$, and $\alpha_h=\alpha_q=2/\beta_h=2/3$.
Then Equation \eqref{eq:corcond} is satisfied if $\rho_{n}=o_p(n^{-5/14})$, while Equation \eqref{eq:corcond2} is satisfied if $\rho_{n}=o_p(n^{-25/64})$. That is, as expected, the requirement of Equation \eqref{eq:corcond2} is slightly stronger. Nevertheless, one can still incorporate non-parametric estimators for nuisance component estimation, as rates slower than the parametric rate, i.e., root-$n$ rate is allowed.

Now, suppose we did not incorporate the debiasing strategy. Then the terms containing $\hat r$ would have vanished, but using a counterpart of the analysis technique for the proof of Theorem \ref{thm:hpc}, we would have had $\|T-\hat T\|$ instead of $\|T-\hat T\|^2$. In this case, the term involving $\|T-\hat T\|$ would dominate and consequently, Equation \eqref{eq:corcond2} is satisfied if $\rho_{n}=o_p(n^{-25/32})$. That is, we would have needed a rate even faster than the parametric rate, which is not possible.

\end{example}

\subsection{Application to Proximal Causal Inference}
\label{sec:proximal}

Proximal causal inference framework is a recently proposed framework for assessing the causal effect of a treatment variable on an outcome variable of interest \citep{miao2018identifying, tchetgen2020introduction}. This framework allows for the presence of an unobserved confounder $U$ in the system, yet, requires having access to two proxy variables of $U$. Let $A\in\{0,1\}$ be a binary treatment variable, $Y\in\mathcal{Y}$ be the outcome variable of interest, and $X\in\mathcal{X}$ be the observed covariates. We use the potential outcomes framework \citep{neyman1923application,rubin1974estimating} and denote the potential outcome variable of $Y$, had the treatment variables been set to value $A=a$ (possibly contrary to the fact) by $Y^{(a)}$, where $a\in\{0,1\}$. We assume the standard consistency and positivity assumptions which require that $Y^{(A)}=Y$, and $0<p(A=1\mid X,U)<1$, respectively. Our causal parameter is the average treatment effect (ATE), defined as $\E[Y^{(1)}-Y^{(0)}]$.

The most common approach for identification of ATE is to assume conditional exchangebility \citep{hernan2020causal}, which requires independence of $Y^{(a)}$ from $A$, for $a\in\{0,1\}$ conditioned on $X$, denoted as $Y^{(a)}\independent A\mid X$. Proximal causal inference framework relaxes this assumption to a version called \emph{latent conditional exchangeability}.
\begin{assumption}[Latent conditional exchangeability]
\label{assm:lex}
For $a\in\{0,1\}$, we have 
\[
Y^{(a)}\independent A\mid X,U.
\]
\end{assumption}
Instead, this framework requires having access to two proxy variables $Z$ and $W$ of the unobserved confounder, which satisfy the following assumption.
\begin{assumption}
\label{assm:prx}
Proxy variables $Z$ and $W$ are directly associated with the unobserved confounder $U$, and satisfy
\begin{align*}
	&Y\independent Z\mid\{A,X,U\},\\
	&W\independent\{A,Z\}\mid \{X,U\}.
\end{align*}
\end{assumption}
Variables $Z$ and $W$ satisfying Assumption \ref{assm:prx} are called the \emph{treatment proxy variable} and the \emph{outcome proxy variable}, respectively.

We focus on the potential outcome mean $\E[Y^{(a)}]$, where $a\in\{0,1\}$. We require the following conditions on the data generating distribution.
\begin{assumption}
\label{assm:proxies}
The proxy variables $Z$ and $W$ satisfy the following.
\begin{enumerate}[label=(\alph*)]

\item There exists an outcome bridge function $h$ that solves the integral equation
\[
\E[I(A=a)h(W,X)\mid Z,A,X]=\E[I(A=a)Y\mid Z,A,X].
\]

\item For square-integrable function $g$ and any $a,x$, if $\E[g(U)\mid Z,a,x]=0$ almost surely, then $g(U)=0$ almost surely.

\item There exists a treatment bridge function $q$ that solves the integral equation
\[
\E[q(Z,A,X)\mid W,A=a,X]=p(A=a\mid W,X)^{-1},
\]
which can be written as 
\[
\E\left[I(A=a)q(Z,A,X)-1\mid W,X\right]=0.
\]

\item For square-integrable function $g$ and any $a,x$, if $\E[g(U)\mid W,a,x]=0$ almost surely, then $g(U)=0$ almost surely.

\end{enumerate}
\end{assumption}
We denote the $L^2$-minimal solutions to integral equations in Parts $(a)$ and $(c)$ of Assumption \ref{assm:proxies} by $h_0$ and $q_0$, respectively.

\cite{miao2018identifying} showed that under Assumptions \ref{assm:lex}, \ref{assm:prx}, \ref{assm:proxies}$(a)$, and \ref{assm:proxies}$(b)$, the potential outcome mean is identified as $\E[Y^{(a)}]=\E[h(W,X)]$.
Later, \cite{cui2023semiparametric} showed that alternatively, under Assumptions \ref{assm:lex}, \ref{assm:prx}, \ref{assm:proxies}$(c)$, and \ref{assm:proxies}$(d)$, the potential outcome mean is identified as $\E[Y^{(a)}]=\E[I(A=a)Yq(W,A,X)]$. Based on these two results, \cite{cui2023semiparametric} obtained the influence function of the parameter 
\[
\psi_0=\E[h(W,X)],
\]
and showed that under the above assumptions, the parameter $\E[Y^{(a)}]$ can be identified by the following functional which incorporates the information of both $h$ and $q$.
\[
\E[Y^{(a)}]=\E\left[-I(A=a)h(W,X)q(W,A,X)+I(A=a)Yq(W,A,X)+h(W,X)\right].
\]
Based on this representation, the following estimator can be used for the parameter $\psi_0$.
\[
\hat\psi=\E_n\left[-I(A=a)\hat h(W,X)\hat q(W,A,X)+I(A=a)Y\hat q(W,A,X)+\hat h(W,X)\right].
\]
where $\hat h$ and $\hat q$ are estimators of $h_0$ and $q_0$, respectively.
Note that this is a special case of the estimator \eqref{eq:estclass} with $V=\{W,Z,A,X,Y\}$, $V_h=\{W,X\}$, $V_q=\{Z,A,X\}$, $s_1(V)=-I(A=a)$, $s_2(V)=I(A=a)Y$, $s_3(V)=1$, and $s_4(V)=0$. Therefore, our results from the previous sections can be directly applied to the proximal causal inference setting.

\section{Conclusion}
\label{sec:conc}

We studied the problem of estimating the $L^2$-minimal solution to a conditional moment equation under smoothness conditions on the operator and the solution. This takes the form of an integral equation, a type of inverse problem that is generally ill-posed. Instances of this problem have been extensively discussed in the literature, especially in the nonparametric instrumental variable framework, where a standard approach is to find the minimizer of a regularized version of the projected mean squared error. However, this approach can be sensitive to misspecification or slow convergence rate of the involved nuisance functions, especially, the estimator of the operator of the integral equation. To address this, we proposed a debiased estimation strategy based on the influence function of a modification of the projected error. We demonstrated that this approach leads to robustness against slow convergence rate of the operator estimator, and obtained finite-sample convergence rate of our proposed estimator using modern statistical learning-theoretic tools to characterize function space complexity. Additionally, we proposed a hyper-parameter tuning strategy and studied the loss in the convergence rate caused by hyper-parameter tuning compared to the case that the optimal order of the hyper-parameter is assumed to be known.  We further studies the application of our approach to the estimation of regular parameters in a specific parameter class, which are linear functionals of the solutions to the conditional moment restrictions. For this class, we established new sufficient conditions for achieving root-$n$ consistency of a debiased influence function-based estimator. This yields an estimation strategy for this setting that incorporates two layers of debiasing.

\section*{Acknowledgements}

Andrea Rotnitzky's work has been supported by the National Heart, Lung, and Blood Institute grant R01-HL137808, and by the National Institute of Allergy and Infectious Diseases grants UM1-AI068635 and R37-AI029168.


\bibliography{Refs.bib}
\bibliographystyle{apalike}

\newpage

\appendices

\section{Proofs}

\subsection*{Proof of Proposition \ref{prop:Yh0}}

The claim follows from observing that
\begin{align*}
\|T(h-h_0)\|_2^2
	=&\E[\E\left[g_1(V)h(V_h)-g_0(V)\mid V_q\right]^2]\\
	=&\E[\{\E[g_1(V)h(V_h)\mid V_q]-g_0(V)+g_0(V)-\E[g_0(V)\mid V_q]\}^2]\\
	=&\E[\{\E[g_1(V)h(V_h)\mid V_q]-g_0(V)\}^2]
	+\E[\{g_0(V)-\E[g_0(V)\mid V_q]\}^2]\\
	&+\E[2\{\E[g_1(V)h(V_h)\mid V_q]-g_0(V)\}\{g_0(V)-\E[g_0(V)\mid V_q]\}]\\
	=&\E[\{\E[g_1(V)h(V_h)\mid V_q]-g_0(V)\}^2]\\
	&-\E[2g_0(V)\{g_0(V)-\E[g_0(V)\mid V_q]\}]
	+\E[\{g_0(V)-\E[g_0(V)\mid V_q]\}^2]\\
	&+\underbrace{\E[2\E[g_1(V)h(V_h)\mid V_q]\{g_0(V)-\E[g_0(V)\mid V_q]\}]}_{=0}.
\end{align*}

\begin{flushright}
$\Box$	
\end{flushright}

\subsection*{Proof of Theorem \ref{thm:IFNN}}

We use the notation $\partial_tg(t)$ to denote $\frac{\partial g(t)}{\partial t}\big|_{t=0}$.
For parameter $\psi(h)$, let ${\psi(h)}_t$ be the parameter of interest under a regular parametric sub-model indexed by $t$, that includes the ground-truth model at $t=0$.
We obtain an influence function of $\psi(h)$ as follows.
We need to find a random variable $M$ with mean zero, that satisfies
\[
\partial_t{\psi(h)}_t=E(MS(V)),
\]
where $S(V)=\partial_t\log p_t(V)$.

Let $G_0=g_0(V)$, and $G_1=g_1(V)$.
Note that
\begin{align*}
\partial_t{\psi(h)}_t
&=\partial_t \E_t[\{\E_t[G_1h(V_h)\mid V_q]-G_0\}^2]\\
&= \partial_t \E_t[\{\E[G_1h(V_h)\mid V_q]-G_0\}^2]\tag{J1}\\
&\quad+ \E[\partial_t\{\E_t[G_1h(V_h)\mid V_q]-G_0\}^2]\tag{J2}.
\end{align*}

For (J1), we have
\begin{align*}
\text{(J1)}
&=\partial_t \E_t[\{\E[G_1h(V_h)\mid V_q]-G_0\}^2]\\
&= \E[\{\E[G_1h(V_h)\mid V_q]-G_0\}^2S(V)]\\
&= \E[\{\{\E[G_1h(V_h)\mid V_q]-G_0\}^2-\psi(h)\}S(V)].
\end{align*}

For (J2), we have
\begin{align*}
\text{(J2)}
&=\E[\partial_t\{\E_t[G_1h(V_h)\mid V_q]-G_0\}^2]\\
&=\E[2\{\E_t[G_1h(V_h)\mid V_q]-G_0\}\partial_t\E_t[G_1h(V_h)\mid V_q]]\\
&=\E[2\{\E_t[G_1h(V_h)\mid V_q]-G_0\}\E[G_1h(V_h)S(V\mid V_q)\mid V_q]]\\
&=\E[2\{\E[G_1h(V_h)\mid V_q]-\E[G_0\mid V_q]\}\E[G_1h(V_h)S(V\mid V_q)\mid V_q]]\\
&=\E[2\{\E[G_1h(V_h)\mid V_q]-\E[G_0\mid V_q]\}\E[\{G_1h(V_h)-\E[G_1h(V_h)\mid V_q]\}S(V\mid V_q)\mid V_q]]\\
&=\E[2\{\E[G_1h(V_h)\mid V_q]-\E[G_0\mid V_q]\}\{G_1h(V_h)-\E[G_1h(V_h)\mid V_q]\}S(V\mid V_q)].
\end{align*}
Note that
\[
\E[2\{\E[G_1h(V_h)\mid V_q]-\E[G_0\mid V_q]\}\{G_1h(V_h)-\E[G_1h(V_h)\mid V_q]\}S(V_q)]=0.
\]
Therefore,
\[
\text{(J2)}=\E[2\{\E[G_1h(V_h)\mid V_q]-\E[G_0\mid V_q]\}\{G_1h(V_h)-\E[G_1h(V_h)\mid V_q]\}S(V)].
\]

Combining the obtained expression for (J1) and (J2), we have
\begin{align*}
\partial_t{\psi(h)}_t
=&\E\big[\big\{\left\{\E[G_1h(V_h)\mid V_q]-G_0\right\}^2\\
&+2\left\{\E[G_1h(V_h)\mid V_q]-\E[G_0\mid V_q]\}\{G_1h(V_h)-\E[G_1h(V_h)\mid V_q]\right\}\\
&-\psi(h)\big\}S(V)\big],
\end{align*}
which concludes the desired result.

\begin{flushright}
$\Box$	
\end{flushright}

\subsection*{Proof of Proposition \ref{prop:biasNN}}

\begingroup
\allowdisplaybreaks
\begin{align*}
&\psi(h)-\E[\hat\psi(h)]\\
&=\E\left[\left\{(Th)(V_q)-g_0(V)\right\}^2-\left\{(\hat Th)(V_q)-g_0(V)\right\}^2-2\left\{(\hat Th)(V_q)-\hat r(V_q)\right\}\left\{g_1(V)h(V_h)-(\hat Th)(V_q)\right\}\right]\\
&=\E[(Th)^2(V_q)-2(Th)(V_q)g_0(V)-(\hat Th)^2(V_q)+2(\hat Th)(V_q)g_0(V)\\
&~~~~~-2(\hat Th)(V_q)g_1(V)h(V_h)+2(\hat Th)^2(V_q)+2\hat r(V_q)g_1(V)h(V_h)-2\hat r(V_q)(\hat Th)(V_q)]\\
&=\E[(Th)^2(V_q)-2(Th)(V_q)r_0(V_q)-(\hat Th)^2(V_q)+2(\hat Th)(V_q)r_0(V_q)\\
&~~~~~-2(\hat Th)(V_q)(Th)(V_q)+2(\hat Th)^2(V_q)+2\hat r(V_q)(Th)(V_q)-2\hat r(V_q)(\hat Th)(V_q)]\\
&=\E\left[\left\{(Th)(V_q)-(\hat Th)(V_q)\right\}^2
+2\left\{(Th)(V_q)-(\hat Th)(V_q)\right\}\left\{\hat r(V_q)-r_0(V_q)\right\}
\right]\\
&\le \|Th-\hat Th\|_2^2+2\|Th-\hat Th\|_2\|r-\hat r_0\|_2.
\end{align*}
\endgroup

\begin{flushright}
$\Box$	
\end{flushright}

\subsection*{Proof of Theorem \ref{thm:mainNN0}}

We first note that
\begingroup
\allowdisplaybreaks
\begin{align*}
\|T(\hath-h_0)\|_2^2
&=\|T(\hath-h_0)\|_2^2-\|T(h_0-h_0)\|_2^2\\
&=\E[\{(T\hath)(V_q)-(Th_0)(V_q)\}^2-\{(Th_0)(V_q)-(Th_0)(V_q)\}^2]\\
&=\E[\{(T\hath)(V_q)-r_0(V_Q)\}^2-\{(Th_0)(V_q)-r_0(V_q)\}^2]\\
&=\E[(T\hath)^2(V_q)+r_0^2(V_Q)-2(T\hath)(V_q)r_0(V_Q)+g_0^2(V)\\
&\qquad-\{(Th_0)^2(V_q)+r_0^2(V_Q)-2(Th_0)(V_q)r_0(V_Q)+g_0^2(V)\}]\\
&=\E[(T\hath)^2(V_q)+g_0^2(V)-2(T\hath)(V_q)g_0(V)\\
&\qquad-\{(Th_0)^2(V_q)+g_0^2(V)-2(Th_0)(V_q)g_0(V)\}]\\
&=\|T\hath-g_0\|_2^2-\|Th_0-g_0\|_2^2,
\end{align*}
\endgroup
where we used the fact that for any $h$,
\begingroup
\allowdisplaybreaks
\begin{align*}
\E[(Th)(V_q)r_0(V_q)]
&=\E[(Th)(V_q)\E[g_0(V)\mid V_q]]\\
&=\E[\E[(Th)(V_q)g_0(V)\mid V_q]]\\
&=\E[(Th)(V_q)g_0(V)].
\end{align*}
\endgroup

By Proposition \ref{prop:biasNN}, we have
\begin{align*}
&\|T\hath-g_0\|_2^2\\
&=\E\left[\left\{(\hat T\hath)(V_q)-g_0(V)\right\}^2+2\left\{(\hat T\hath)(V_q)-\hat r(V_q)\right\}\left\{g_1(V)\hath(V_h)-(\hat T\hath)(V_q)\right\}\right]\\
&~~~+\|T\hath-\hat T\hath\|_2^2+2\E\left[\left\{(T\hath)(V_q)-(\hat T\hath)(V_q)\right\}\left\{\hat r(V_q)-r_0(V_q)\right\}\right].
\end{align*}
Similarly, 
\begin{align*}
&\|Th_0-g_0\|_2^2\\
&=\E\left[\left\{(\hat Th_0)(V_q)-g_0(V)\right\}^2+2\left\{(\hat Th_0)(V_q)-\hat r(V_q)\right\}\left\{g_1(V)h_0(V_h)-(\hat Th_0)(V_q)\right\}\right]\\
&~~~+\|Th_0-\hat Th_0\|_2^2+2\E\left[\left\{(Th_0)(V_q)-(\hat Th_0)(V_q)\right\}\left\{\hat r(V_q)-r_0(V_q)\right\}\right].
\end{align*}
Therefore,
\begin{equation}
\label{eq:pt01}
\|T(\hath-h_0)\|_2^2
=\|T\hath-g_0\|_2^2
-\|T h_0-g_0\|_2^2
=(\text{J}1)+(\text{J}2),
\end{equation}
where,
\begin{align*}
(\text{J}1)
&=
\E\left[\left\{(\hat T\hath)(V_q)-g_0(V)\right\}^2+2\left\{(\hat T\hath)(V_q)-\hat r(V_q)\right\}\left\{g_1(V)\hath(V_h)-(\hat T\hath)(V_q)\right\}\right]\\
&~~~-
\E\left[\left\{(\hat Th_0)(V_q)-g_0(V)\right\}^2+2\left\{(\hat Th_0)(V_q)-\hat r(V_q)\right\}\left\{g_1(V)h_0(V_h)-(\hat Th_0)(V_q)\right\}\right],
\end{align*}
and
\begin{equation}
\label{eq:pt02}
\begin{aligned}
(\text{J}2)
&=
\|T\hath-\hat T\hath\|_2^2-\|Th_0-\hat Th_0\|_2^2\\
&~~~+2\E\left[\left\{(T\hath)(V_q)-(\hat T\hath)(V_q)\right\}\left\{\hat r(V_q)-r_0(V_q)\right\}\right]\\
&~~~-2\E\left[\left\{(Th_0)(V_q)-(\hat Th_0)(V_q)\right\}\left\{\hat r(V_q)-r_0(V_q)\right\}\right]\\
&=\E\left[((T-\hat T)(\hath-h_0))(V_q)((T-\hat T)(\hath+h_0))(V_q)\right]\\
&~~~+2\E\left[((T-\hat T)(\hath-h_0))(V_q)\left\{\hat r(V_q)-r_0(V_q)\right\}\right]\\
&\le \|(T-\hat T)(\hath-h_0)\|_2\|(T-\hat T)(\hath+h_0)\|_2\\
&~~~+2\|(T-\hat T)(\hath-h_0)\|_2\|\hat r-r_0\|_2\\
&\lesssim \|\hat T-T\|^2\|\hath-h_0\|_2+2\|\hat T-T\|\|\hat r-r_0\|_2\|\hath-h_0\|_2.
\end{aligned}
\end{equation}

Regarding $(\text{J}1)$, We define the loss function
\begin{align*}
\ell(V;h)
&:=\left\{(\hat Th)(V_q)-g_0(V)\right\}^2+2\left\{(\hat Th)(V_q)- \hat  r(V_q)\right\}\left\{g_1(V)h(V_h)-(\hat Th)(V_q)\right\}\\
&\quad+\lambda\{h(V_h)-\hat h_{\lambda,t-1}^\text{IF}(V_h)\}^2.
\end{align*}
Then, we have
\begingroup
\allowdisplaybreaks
\begin{align*}
(\text{J}1)
&=\E[\ell(V;\hath)-\ell(V;h_0)]
-\lambda\|\hath-\hat h_{\lambda,t-1}^\text{IF}\|^2_2+\lambda\|h_0-\hat h_{\lambda,t-1}^\text{IF}\|^2_2\\
&\lesssim\E[\ell(V;\hath)-\ell(V;h_0)]+\lambda\|\hath-h_0\|_2\\
&=\E_n[\ell(V;\hath)-\ell(V;h_0)]\\
&\qquad+(\E-\E_n)[\ell(V;\hath)-\ell(V;h_0)]+\lambda\|\hath-h_0\|_2\\
&\le(\E-\E_n)[\ell(V;\hath)-\ell(V;h_0)]+\lambda\|\hath-h_0\|_2,
\end{align*}
\endgroup
where the last inequality follows from the definition of $\hath$ and that $h_0$ is assumed to belong to $\mathcal{H}$.

Note that
\begingroup
\allowdisplaybreaks
\begin{align*}
&(\E-\E_n)[\ell(V;\hath)-\ell(V;h_0)]\\
&=(\E-\E_n)[\{(\hat T\hath)(V_q)-g_0(V)\}^2-\{(\hat Th_0)(V_q)-g_0(V)\}^2]\\
&\qquad+(\E-\E_n)[2\{(\hat T\hath)(V_q)-(\hat Th_0)(V_q)+(\hat Th_0)(V_q)-(Th_0)(V_q)+(Th_0)(V_q)-\hat r(V_q)\}g_1(V)\hath(V_h)]\\
&\qquad+(\E-\E_n)[-2\{(\hat Th_0)(V_q)-(Th_0)(V_q)+(Th_0)(V_q)-\hat r(V_q)\}g_1(V)h_0(V_h)]\\
&\qquad+(\E-\E_n)[-2\{(\hat T\hath)(V_q)-(\hat Th_0)(V_q)+(\hat Th_0)(V_q)-(Th_0)(V_q)+(Th_0)(V_q)-\hat r(V_q)\}(\hat T\hath)(V_q)]\\
&\qquad+(\E-\E_n)[2\{(\hat Th_0)(V_q)-(Th_0)(V_q)+(Th_0)(V_q)-\hat r(V_q)\}(\hat Th_0)(V_q)]\\
&\qquad+\lambda(\E-\E_n)[\{\hath(V_h)-\hat h_{\lambda,t-1}^\text{IF}(V_h)\}^2-\{h_0(V_h)-\hat h_{\lambda,t-1}^\text{IF}(V_h)\}^2]\\
&=(\E-\E_n)[\{(\hat T\hath)(V_q)-(\hat Th_0)(V_q)\}\{(\hat T\hath)(V_q)+(\hat Th_0)(V_q)-2g_0(V)\}]\\
&\qquad+(\E-\E_n)[2\{(\hat T\hath)(V_q)-(\hat Th_0)(V_q)\}\{g_1(V)\hath(V_h)-(\hat T\hath)(V_q)\}]\\
&\qquad+(\E-\E_n)[2\{(\hat Th_0)(V_q)-(Th_0)(V_q)\}\{g_1(V)\hath(V_h)-g_1(V)h_0(V_h)-(\hat T\hath)(V_q)+(\hat Th_0)(V_q)\}]\\
&\qquad+(\E-\E_n)[2\{r_0(V_q)-\hat r(V_q)\}\{g_1(V)\hath(V_h)-g_1(V)h_0(V_h)-(\hat T\hath)(V_q)+(\hat Th_0)(V_q)\}]\\
&\qquad+\lambda(\E-\E_n)[\{\hath(V_h)-h_0(V_h)\}\{\hath(V_h)+h_0(V_h)-2\hat h_{\lambda,t-1}^\text{IF}(V_h)]\\
&\lesssim|(\E-\E_n)[(\hat T\hath)(V_q)-(\hat Th_0)(V_q)]|\\
&\qquad+|(\E-\E_n)[(\hat T\hath)(V_q)-(\hat Th_0)(V_q)]|\\
&\qquad+|(\E-\E_n)[(\hat Th_0)(V_q)-(Th_0)(V_q)]|\\
&\qquad+|(\E-\E_n)[r_0(V_q)-\hat r(V_q)]|\\
&\qquad+\lambda|(\E-\E_n)[\hath(V_h)-h_0(V_h)]|\\
&\lesssim\delta_{M,n}\|\hat T(\hath-h_0)\|_2+\delta_{M,n}^2\\
&\qquad+\delta_{M,n}\|(\hat T-T)h_0\|_2+\delta_{M,n}^2\\
&\qquad+\delta_{M,n}\|\hat r-r_0\|_2+\delta_{M,n}^2\\
&\qquad+\lambda\delta_{n}\|\hath-h_0\|_2+\lambda\delta_{n}^2,
\end{align*}
\endgroup
with probability at least $1-5\zeta$, where the last inequality is due to Lemma \ref{lem:cons}.

Therefore, with probability at least $1-5\zeta$, we have
\begingroup
\allowdisplaybreaks
\begin{align*}
&(\text{J}1)\\
&\lesssim
\delta_{M,n}\{\|\hat T(\hath-h_0)\|_2+\|(\hat T-T)h_0\|_2+\|\hat r-r_0\|_2\}+\delta_{M,n}^2+\lambda\delta_{n}\|\hath-h_0\|_2+\lambda\delta_{n}^2+\lambda\|\hath-h_0\|_2\\
&\lesssim
\delta_{M,n}\{\|\hat T(\hath-h_0)\|_2+\|\hat T-T\|+\|\hat r-r_0\|_2\}+\delta_{M,n}^2+\lambda\delta_{n}^2+\lambda\|\hath-h_0\|_2.
\end{align*}
\endgroup
Note that
\begingroup
\allowdisplaybreaks
\begin{align*}
\|\hat T(\hath-h_0)\|_2
&\le \|(\hat T-T)(\hath-h_0)\|_2+\|T(\hath-h_0)\|_2\\
&\le \|\hat T-T\|+\|T(\hath-h_0)\|_2.
\end{align*}
\endgroup
Therefore, with probability at least $1-5\zeta$, we have
\begin{equation}
\label{eq:pt03}
(\text{J}1)
\lesssim
\delta_{M,n}\{\|T(\hath-h_0)\|_2+\|\hat T-T\|+\|\hat r-r_0\|_2\}+\delta_{M,n}^2+\lambda\delta_{n}^2+\lambda\|\hath-h_0\|_2.
\end{equation}

Combining \eqref{eq:pt01}, \eqref{eq:pt02}, and \eqref{eq:pt03}, we conclude that, with probability at least $1-5\zeta$, we have
\begingroup
\allowdisplaybreaks
\begin{align*}
\|T(\hath-h_0)\|_2^2
\lesssim
&\{\|T(\hath-h_0)\|_2+\|\hat T-T\|+\|\hat r-r_0\|_2\}\delta_{M,n}\\
&+\|\hat T-T\|^2\|\hath-h_0\|_2+\|\hat T-T\|\|\hat r-r_0\|_2\|\hath-h_0\|_2\\
&+\delta_{M,n}^2+\lambda\delta_{n}^2+\lambda\|\hath-h_0\|_2.
\end{align*}
\endgroup
Using the weighted AM-GM inequality,
\[
ab\le \frac{a^2}{2w}+\frac{wb^2}{2},
\]
we have
\begingroup
\allowdisplaybreaks
\begin{align*}
&\{\|T(\hath-h_0)\|_2+\|\hat T-T\|+\|\hat r-r_0\|_2\}\delta_{M,n}\\
&\le \frac{1}{2w}\{\|T(\hath-h_0)\|_2+\|\hat T-T\|+\|\hat r-r_0\|_2\}^2
+\frac{w}{2}\delta_{M,n}^2\\
&\le \frac{2}{w}\|T(\hath-h_0)\|_2^2+\frac{2}{w}\|\hat T-T\|^2+\frac{2}{w}\|\hat r-r_0\|_2^2
+\frac{w}{2}\delta_{M,n}^2.
\end{align*}
\endgroup
Therefore, there exists positive constants $c_1$ and $c_2$, with $c_1<1$, such that, with probability at least $1-5\zeta$, we have
\begingroup
\allowdisplaybreaks
\begin{align*}
\|T(\hath-h_0)\|_2^2
\lesssim~
&c_1\|T(\hath-h_0)\|_2^2\\
+&c_2\Big\{\|\hat T-T\|^2+\|\hat r-r_0\|_2^2
+\delta_{M,n}^2+\lambda\delta_{n}^2+\lambda\|\hath-h_0\|_2\Big\},
\end{align*}
\endgroup
which implies that
\begingroup
\allowdisplaybreaks
\begin{align*}
(1-c_1)\|T(\hath-h_0)\|_2^2
\lesssim
\|\hat T-T\|^2+\|\hat r-r_0\|_2^2
+\delta_{M,n}^2+\lambda\delta_{n}^2+\lambda\|\hath-h_0\|_2,
\end{align*}
\endgroup
or equivalently,
\begingroup
\allowdisplaybreaks
\begin{align*}
\|T(\hath-h_0)\|_2^2
\lesssim
\|\hat T-T\|^2+\|\hat r-r_0\|_2^2
+\delta_{M,n}^2+\lambda\delta_{n}^2+\lambda\|\hath-h_0\|_2,
\end{align*}
\endgroup
with probability at least $1-5\zeta$.

\begin{flushright}
$\Box$	
\end{flushright}

\subsection*{Proof of Theorem \ref{thm:mainNN}}

Recall the definition of the regularized population-level objective function $\starh$ in Display \eqref{eq:popobjNN}.
By triangle inequality, we have
\begin{equation}
\label{eq:mpf1NN}
\|\hath-h_0\|_2^2
\le
2\|\hath-\starh\|_2^2
+
2\|\starh-h_0\|_2^2.
\end{equation}
Here, the second term in the upper bound characterizes the regularization bias.
By Lemma \ref{lem:reg}, under Assumption \ref{assm:beta}, this term can be bounded as
\begin{equation}
\label{eq:mpf2NN}
\|\starh-h_0\|_2^2\le B\lambda^{\min\{2t,\beta\}}.
\end{equation}
Hence, it suffices to bound $\|\hath-\starh\|_2^2$.

Define
\[
L(\tau):=
\left\|T\left(\starh+\tau(\hath-\starh)\right)-g_0(V)\right\|_2^2
+\lambda\left\|\starh+\tau(\hath-\starh)-h^*_{\lambda,t-1}\right\|_2^2.
\]
Similar to \citep{bennett2023source}, we start by leveraging the convexity of $L$. By the definition of $\starh$, $L(\tau)$ is minimized at $\tau=0$, and hence, $\frac{d}{d\tau}L(\tau)|_{\tau=0}=0$. Therefore, by Taylor expansion, we have
\begin{align*}
L(1)-L(0)
&=\frac{d}{d\tau}L(\tau)|_{\tau=0}+\frac{1}{2}\frac{d^2}{d\tau^2}L(\tau)|_{\tau=0}\\
&=\frac{1}{2}\frac{d^2}{d\tau^2}L(\tau)|_{\tau=0}.
\end{align*}
Therefore, 
\begin{align*}
&\lambda\|\hath-\starh\|_2^2+\|T(\hath-\starh)\|_2^2\\
&=
\{\|T\hath-g_0(V)\|_2^2+\lambda\|\hath-h^*_{\lambda,t-1}\|_2^2\}
-\{\|T \starh-g_0(V)\|_2^2+\lambda\|\starh-h^*_{\lambda,t-1}\|_2^2\}.
\end{align*}

By Proposition \ref{prop:biasNN}, we have
\begin{align*}
&\|T\hath-g_0(V)\|_2^2\\
&=\E\left[\left\{(\hat T\hath)(V_q)-g_0(V)\right\}^2+2\left\{(\hat T\hath)(V_q)-\hat r(V_q)\right\}\left\{g_1(V)\hath(V_h)-(\hat T\hath)(V_q)\right\}\right]\\
&~~~+\|T\hath-\hat T\hath\|_2^2+2\E\left[\left\{(T\hath)(V_q)-(\hat T\hath)(V_q)\right\}\left\{\hat r(V_q)-r_0(V_q)\right\}\right].
\end{align*}
Similarly, 
\begin{align*}
&\|T\starh-g_0(V)\|_2^2\\
&=\E\left[\left\{(\hat T\starh)(V_q)-g_0(V)\right\}^2+2\left\{(\hat T\starh)(V_q)-\hat r(V_q)\right\}\left\{g_1(V)\starh(V_h)-(\hat T\starh)(V_q)\right\}\right]\\
&~~~+\|T\starh-\hat T\starh\|_2^2+2\E\left[\left\{(T\starh)(V_q)-(\hat T\starh)(V_q)\right\}\left\{\hat r(V_q)-r_0(V_q)\right\}\right].
\end{align*}
Therefore,
\begin{align*}
\{\|T\hath-g_0(V)\|_2^2+\lambda\|\hath-h^*_{\lambda,t-1}\|_2^2\}
-\{\|T \starh-g_0(V)\|_2^2+\lambda\|\starh-h^*_{\lambda,t-1}\|_2^2\}
=(\text{J}1)+(\text{J}2),
\end{align*}
where,
\begin{align*}
&(\text{J}1)\\
&=
\E\left[\left\{(\hat T\hath)(V_q)-g_0(V)\right\}^2+2\left\{(\hat T\hath)(V_q)-\hat r(V_q)\right\}\left\{g_1(V)\hath(V_h)-(\hat T\hath)(V_q)\right\}\right]\\
&~~~+\lambda\E\left[\{\hath(V_h)-h^*_{\lambda,t-1}(V_h)\}^2\right]\\
&~~~-
\E\left[\left\{(\hat T\starh)(V_q)-g_0(V)\right\}^2+2\left\{(\hat T\starh)(V_q)-\hat r(V_q)\right\}\left\{g_1(V)\starh(V_h)-(\hat T\starh)(V_q)\right\}\right]\\
&~~~-\lambda\E\left[\{\starh(V_h)-h^*_{\lambda,t-1}(V_h)\}^2\right],
\end{align*}
and
\begin{align*}
&(\text{J}2)\\
&=
\|T\hath-\hat T\hath\|_2^2-\|T\starh-\hat T\starh\|_2^2\\
&~~~+2\E\left[\left\{(T\hath)(V_q)-(\hat T\hath)(V_q)\right\}\left\{\hat r(V_q)-r_0(V_q)\right\}\right]\\
&~~~-2\E\left[\left\{(T\starh)(V_q)-(\hat T\starh)(V_q)\right\}\left\{\hat r(V_q)-r_0(V_q)\right\}\right].
\end{align*}

Regarding $(\text{J}1)$, We define the loss function
\begin{align*}
\ell(V;h)
&:=\left\{(\hat Th)(V_q)-g_0(V)\right\}^2+2\left\{(\hat Th)(V_q)- \hat  r(V_q)\right\}\left\{g_1(V)h(V_h)-(\hat Th)(V_q)\right\}\\
&\quad+\lambda\{h(V_h)-\hat h_{\lambda,t-1}^\text{IF}(V_h)\}^2.
\end{align*}
Then, we have
\begingroup
\allowdisplaybreaks
\begin{align*}
(\text{J}1)
&=\E[\ell(V;\hath)-\ell(V;\starh)]\\
&\quad+\{\lambda\|\hath-h^*_{\lambda,t-1}\|_2^2-\lambda\|\starh-h^*_{\lambda,t-1}\|_2^2\}\\
&\quad-\{\lambda\|\hath-\hat h_{\lambda,t-1}^\text{IF}\|_2^2-\lambda\|\starh-\hat h_{\lambda,t-1}^\text{IF}\|_2^2\}\\
&=(\E-\E_n)[\ell(V;\hath)-\ell(V;\starh)]\\
&\quad+\E_n[\ell(V;\hath)]-\E_n[\ell(V;\starh)]\\
&\quad+\{\lambda\|\hath-h^*_{\lambda,t-1}\|_2^2-\lambda\|\starh-h^*_{\lambda,t-1}\|_2^2\}\\
&\quad-\{\lambda\|\hath-\hat h_{\lambda,t-1}^\text{IF}\|_2^2-\lambda\|\starh-\hat h_{\lambda,t-1}^\text{IF}\|_2^2\}\\
&\le(\E-\E_n)[\ell(V;\hath)-\ell(V;\starh)]\\
&\quad+\{\lambda\|\hath-h^*_{\lambda,t-1}\|_2^2-\lambda\|\starh-h^*_{\lambda,t-1}\|_2^2\}\\
&\quad-\{\lambda\|\hath-\hat h_{\lambda,t-1}^\text{IF}\|_2^2-\lambda\|\starh-\hat h_{\lambda,t-1}^\text{IF}\|_2^2\},
\end{align*}
\endgroup
where the inequality follows from the definition of $\hath$.

We note that by Assumption \ref{assm:bdd}, $\ell$ is Lipschitz in its argument $h$. Specifically, we have
\begingroup
\allowdisplaybreaks
\begin{align*}
\|\ell(V;h_1)-\ell(V;h_2)\|_2
&=
\|\{(\hat Th_1)(V_q)-g_0(V)\}^2-\{(\hat Th_2)(V_q)-g_0(V)\}^2\\
&~~~+2\{(\hat Th_1)(V_q)- \hat  r(V_q)\}\{g_1(V)h_1(V_h)-(\hat Th_1)(V_q)\}\\
&~~~-2\{(\hat Th_2)(V_q)- \hat  r(V_q)\}\{g_1(V)h_2(V_h)-(\hat Th_2)(V_q)\}\\
&~~~+\lambda\{h_1^2(V_h)-h_2^2(V_h)\}-2\lambda\hat h_{\lambda,t-1}^\text{IF}\{h_1(V_h)-h_2(V_h)\}\|_2\\
&=
\|(\hat Th_2)^2(V_q)-(\hat Th_1)^2(V_q)\\
&~~~+2\hat r(V_q)g_1(V)\{h_2(V_h)-h_1(V_h)\}\\
&~~~+2(\hat r(V_q)-g_0(V))(\hat T(h_1-h_2))(V_q)\\
&~~~+2g_1(V)\{(\hat Th_1)(V_q)h_1(V_h)-(\hat Th_2)(V_q)h_2(V_h)\}\\
&~~~+\lambda\{h_1^2(V_h)-h_2^2(V_h)\}-2\lambda h^*_{\lambda,t-1}\{h_1(V_h)-h_2(V_h)\}\\
&~~~+2\lambda\{h_{\lambda,t-1}^*-\hat h_{\lambda,t-1}^\text{IF}\}\{h_1(V_h)-h_2(V_h)\}\|_2\\
&\le 
\|(\hat Th_2)^2(V_q)-(\hat Th_1)^2(V_q)\|_2\\
&~~~+\|2\hat r(V_q)g_1(V)\{h_2(V_h)-h_1(V_h)\}\|_2\\
&~~~+\|2(\hat r(V_q)-g_0(V))(\hat T(h_1-h_2))(V_q)\|_2\\
&~~~+\|2g_1(V)\{(\hat Th_1)(V_q)h_1(V_h)-(\hat Th_1)(V_q)h_2(V_h)\}\|_2\\
&~~~+\|2g_1(V)\{(\hat Th_1)(V_q)h_2(V_h)-(\hat Th_2)(V_q)h_2(V_h)\}\|_2\\
&~~~+\|\lambda\{h_1^2(V_h)-h_2^2(V_h)\}\|_2\\
&~~~+\|2\lambda h^*_{\lambda,t-1}\{h_1(V_h)-h_2(V_h)\}\|_2\\
&~~~+\|2\lambda\{h_{\lambda,t-1}^*-\hat h_{\lambda,t-1}^\text{IF}\}\{h_1(V_h)-h_2(V_h)\}\|_2\\
&\le
\|(\hat T(h_1+h_2))(V_q)(\hat T(h_2-h_1))(V_q)\|_2\\
&~~~+\|2\hat r(V_q)g_1(V)\{h_2(V_h)-h_1(V_h)\}\|_2\\
&~~~+\|2(\hat r(V_q)-g_0(V))(\hat T(h_1-h_2))(V_q)\|_2\\
&~~~+\|2g_1(V)(\hat Th_1)(V_q)\{h_1(V_h)-h_2(V_h)\}\|_2\\
&~~~+\|2g_1(V)h_2(V_h)(\hat T(h_1-h_2))(V_q)\|_2\\
&~~~+\|\lambda\{h_1(V_h)+h_2(V_h)\}\{h_1(V_h)-h_2(V_h)\}\|_2\\
&~~~+\|2\lambda h^*_{\lambda,t-1}\{h_1(V_h)-h_2(V_h)\}\|_2\\
&~~~+\|2\lambda\{h_{\lambda,t-1}^*-\hat h_{\lambda,t-1}^\text{IF}\}\{h_1(V_h)-h_2(V_h)\}\|_2\\
&\lesssim
\|h_1-h_2\|_2+\|\hat T(h_1-h_2)\|_2\\
&\le \|h_1-h_2\|_2+\|(\hat T-T)(h_1-h_2)\|_2+\|T(h_1-h_2)\|_2\\
&\le \|h_1-h_2\|_2+\|(\hat T-T)\|\|h_1-h_2\|_2+\|T(h_1-h_2)\|_2\\
&\le \|h_1-h_2\|_2+\|T(h_1-h_2)\|_2\\
&\lesssim
\|h_1-h_2\|_2,
\end{align*}
\endgroup
where in the last step, we used the fact that $T$ is a contraction.

Therefore, by Lemma \ref{lem:cons}, with probability at least $1-\zeta$, we have
\begin{equation}
\label{eq:1J1}
\Big|(\E-\E_n)\left[\ell(V;\hath)-\ell(V;\starh)\right]\Big|\lesssim\delta_n\|\hath-\starh\|_2+\delta_n^2.
\end{equation}

Moreover, we note that
\begingroup
\allowdisplaybreaks
\begin{align*}
&\{\lambda\|\hath-h^*_{\lambda,t-1}\|_2^2-\lambda\|\starh-h^*_{\lambda,t-1}\|_2^2\}\\
&\quad-\{\lambda\|\hath-\hat h_{\lambda,t-1}^\text{IF}\|_2^2-\lambda\|\starh-\hat h_{\lambda,t-1}^\text{IF}\|_2^2\}\\
&\le2\lambda\|\hat h_{\lambda,t-1}^\text{IF}-h^*_{\lambda,t-1}\|_2\|\hath-\starh\|_2.
\end{align*}
\endgroup

This together with Displays \eqref{eq:1J1} imply that, with probability at least $1-\zeta$, we have
\begin{align*}
&(\text{J}1)
\lesssim\delta_n\|\hath-\starh\|_2+2\lambda\|\hat h_{\lambda,t-1}^\text{IF}-h^*_{\lambda,t-1}\|_2\|\hath-\starh\|_2+\delta_n^2.
\end{align*}

Regarding $(\text{J}2)$, we have
\begingroup
\allowdisplaybreaks
\begin{align*}
(\text{J}2)
&=
\|T\hath-\hat T\hath\|_2^2-\|T\starh-\hat T\starh\|_2^2\\
&~~~+2\E\left[\left\{(T\hath)(V_q)-(\hat T\hath)(V_q)\right\}\left\{\hat r(V_q)-r_0(V_q)\right\}\right]\\
&~~~-2\E\left[\left\{(T\starh)(V_q)-(\hat T\starh)(V_q)\right\}\left\{\hat r(V_q)-r_0(V_q)\right\}\right]\\
&=\E\Big[\left\{\left((T-\hat T)\hath\right)(V_q)\right\}^2-\left\{\left((T-\hat T)\starh\right)(V_q)\right\}^2\\
&~~~+2\left\{\left((T-\hat T)(\hath-\starh)\right)(V_q)\right\}\left\{\hat r(V_q)-r_0(V_q)\right\}\Big]\\
&=\E\Big[\left\{\left((T-\hat T)(\hath-\starh)\right)(V_q)\right\} \left\{\left((T-\hat T)(\hath+\starh)\right)(V_q)\right\}\\
&~~~+2\left\{\left((T-\hat T)(\hath-\starh)\right)(V_q)\right\}\left\{\hat r(V_q)-r_0(V_q)\right\}\Big]\\
&=\E\Big[\left\{\left((T-\hat T)(\hath-\starh)\right)(V_q)\right\} \left\{\left((T-\hat T)(\hath-\starh)\right)(V_q)\right\}\\
&~~~+\E\Big[\left\{\left((T-\hat T)(\hath-\starh)\right)(V_q)\right\} \left\{\left((T-\hat T)(2\starh)\right)(V_q)\right\}\\
&~~~+2\left\{\left((T-\hat T)(\hath-\starh)\right)(V_q)\right\}\left\{\hat r(V_q)-r_0(V_q)\right\}\Big]\\
&\le 
\left\|(T-\hat T)(\hath-\starh)\right\|_2^2\\
&~~~+2\left\|(T-\hat T)(\hath-\starh)\right\|_2
\left\|(T-\hat T)(\starh)\right\|_2\\
&~~~+2\left\|(T-\hat T)(\hath-\starh)\right\|_2
\left\|\hat r-r_0\right\|_2\\
&\lesssim 
+\|T-\hat T\|^2\|\hath-\starh\|_2
+\|T-\hat T\|\|\hat r-r_0\|_2\|\hath-\starh\|_2,
\end{align*}
\endgroup
where we used Assumption \ref{assm:bdd} to obtain the last expression.

Combining the bounds on $(\text{J}1)$ and $(\text{J}2)$, we obtain
\begin{align*}
&\lambda\|\hath-\starh\|_2^2+\|T(\hath-\starh)\|_2^2\\
&\lesssim 
\left\{
\delta_n+\|T-\hat T\|^2+\|T-\hat T\|\|\hat r-r_0\|_2
+2\lambda\|\hat h_{\lambda,t-1}^\text{IF}-h^*_{\lambda,t-1}\|_2
\right\}\|\hath-\starh\|_2+\delta_n^2.
\end{align*}

Using the weighted AM-GM inequality,
\[
ab\le \frac{a^2}{2w}+\frac{wb^2}{2},
\]
we have
\begin{align*}
&\lambda\|\hath-\starh\|_2^2+\|T(\hath-\starh)\|_2^2\\
&\le \frac{c_1}{\lambda}
\left\{
\delta_n+\|T-\hat T\|^2+\|T-\hat T\|\|\hat r-r_0\|_2
+2\lambda\|\hat h_{\lambda,t-1}^\text{IF}-h^*_{\lambda,t-1}\|_2
\right\}^2+
c_2\lambda\|\hath-\starh\|_2^2+c_3\delta_n^2,
\end{align*}
for some positive constants $c_1$, $c_2$, and $c_3$, where the weight in the weighted AM-GM inequality is chosen such that $c_2<1$. Therefore, 
\begin{equation}
\label{eq:asl}
\begin{aligned}
&(1-c_2)\lambda\|\hath-\starh\|_2^2+\|T(\hath-\starh)\|_2^2\\
&\le \frac{c_1}{\lambda}
\left\{
\delta_n+\|T-\hat T\|^2+\|T-\hat T\|\|\hat r-r_0\|_2
+2\lambda\|\hat h_{\lambda,t-1}^\text{IF}-h^*_{\lambda,t-1}\|_2
\right\}^2+
c_3\delta_n^2.	
\end{aligned}
\end{equation}

Regarding bounding the source error, Equation \eqref{eq:asl} implies
\begin{align*}
\|\hath-\starh\|_2^2
&\lesssim 
\frac{1}{\lambda^2}
\left\{\delta_n^2+\|T-\hat T\|^4+\|T-\hat T\|^2\|\hat r-r_0\|_2^2\right\}
+\|\hat h_{\lambda,t-1}^\text{IF}-h^*_{\lambda,t-1}\|_2^2\\
&\lesssim 
\frac{1}{\lambda^2}
\max
\left\{\delta_n^2,\|T-\hat T\|^4,\|T-\hat T\|^2\|\hat r-r_0\|_2^2\right\}
+\|\hat h_{\lambda,t-1}^\text{IF}-h^*_{\lambda,t-1}\|_2^2.
\end{align*}
Recall that $\hat h_{\lambda,0}^\text{IF}=h^*_{\lambda,0}=0$.
Therefore, for $t=1$, we have
\begin{equation}
\label{eq:concp1}
\|\hat h_{\lambda,1}^\text{IF}-h^*_{\lambda,1}\|_2^2
\lesssim 
\frac{1}{\lambda^2}
\left\{\delta_n^2+\|T-\hat T\|^4+\|T-\hat T\|^2\|\hat r-r_0\|_2^2\right\}.	
\end{equation}
Therefore,
\begin{equation}
\label{eq:mpf3NN}
\begin{aligned}
\|\hat h_{\lambda,2}^\text{IF}-h^*_{\lambda,2}\|_2^2
&\lesssim 
\frac{1}{\lambda^2}
\left\{\delta_n^2+\|T-\hat T\|^4+\|T-\hat T\|^2\|\hat r-r_0\|_2^2\right\}
+\|\hat h_{\lambda,1}^\text{IF}-h^*_{\lambda,1}\|_2^2\\
&\lesssim 
\frac{1}{\lambda^2}
\left\{\delta_n^2+\|T-\hat T\|^4+\|T-\hat T\|^2\|\hat r-r_0\|_2^2\right\}.
\end{aligned}
\end{equation}

Equations \eqref{eq:mpf1NN}, \eqref{eq:mpf2NN}, and \eqref{eq:mpf3NN} conclude that
\begin{align*}
\|\hat h_{\lambda,2}^\text{IF}-h_0\|_2^2
\lesssim	
\frac{1}{\lambda^2}
\max
\left\{\delta_n^2,\|T-\hat T\|^4,\|T-\hat T\|^2\|\hat r-r_0\|_2^2\right\}
+
\lambda^{\min\{4,\beta\}}.
\end{align*}

Regarding bounding the projected error, by triangle inequality, we have
\begin{equation}
\label{eq:mpf1ProjNN}
\|T(\hath-h_0)\|_2^2
\le
2\|T(\hath-\starh)\|_2^2
+
2\|T(\starh-h_0)\|_2^2.
\end{equation}
Here, the second term in the upper bound characterizes the regularization bias.
By Lemma \ref{lem:reg} for projected error, under Assumption \ref{assm:beta}, this term can be bounded as
\begin{equation}
\label{eq:mpf2ProjNN}
\|T(\starh-h_0)\|_2^2\le B\lambda^{\min\{2t,\beta+1\}}.
\end{equation}
Hence, it suffices to bound $\|T(\hath-\starh)\|_2^2$.

Using Equation \eqref{eq:asl}, we have
\begin{equation}
\label{eq:mpf3ProjNN}
\begin{aligned}
\|T(\hat h_{\lambda,2}^\text{IF}-h^*_{\lambda,2})\|_2^2
&\lesssim 
\frac{1}{\lambda}
\left\{\delta_n^2+\|T-\hat T\|^4+\|T-\hat T\|^2\|\hat r-r_0\|_2^2\right\}
+\lambda\|\hat h_{\lambda,1}^\text{IF}-h^*_{\lambda,1}\|_2^2\\
&\lesssim 
\frac{1}{\lambda}
\left\{\delta_n^2+\|T-\hat T\|^4+\|T-\hat T\|^2\|\hat r-r_0\|_2^2\right\}\\
&\lesssim 
\frac{1}{\lambda}
\max
\left\{\delta_n^2,\|T-\hat T\|^4,\|T-\hat T\|^2\|\hat r-r_0\|_2^2\right\},
\end{aligned}
\end{equation}
where we used Display \eqref{eq:concp1}.
Equations \eqref{eq:mpf1ProjNN}, \eqref{eq:mpf2ProjNN}, and \eqref{eq:mpf3ProjNN} conclude that
\begin{align*}
\|T(\hat h_{\lambda,2}^\text{IF}-h_0)\|_2^2
\lesssim	
\frac{1}{\lambda}
\max
\left\{\delta_n^2,\|T-\hat T\|^4,\|T-\hat T\|^2\|\hat r-r_0\|_2^2\right\}
+
\lambda^{\min\{4,\beta+1\}}.
\end{align*}

\begin{flushright}
$\Box$	
\end{flushright}

\subsection*{Proof of Theorem \ref{thm:mainNN1}}

With an approach similar to that in the proof of Theorem \ref{thm:mainNN}, we have
\begin{align*}
&\lambda\|\hath-\starh\|_2^2+\|T(\hath-\starh)\|_2^2\\
&=
\{\|T\hath-g_0(V)\|_2^2+\lambda\|\hath-h^*_{\lambda,t-1}\|_2^2\}
-\{\|T \starh-g_0(V)\|_2^2+\lambda\|\starh-h^*_{\lambda,t-1}\|_2^2\}\\
&=(\text{J}1)+(\text{J}2),
\end{align*}
where,
\begingroup
\allowdisplaybreaks
\begin{align*}
(\text{J}1)
&=\E[\ell(V;\hath)-\ell(V;\starh)]\\
&\quad+\{\lambda\|\hath-h^*_{\lambda,t-1}\|_2^2-\lambda\|\starh-h^*_{\lambda,t-1}\|_2^2\}\\
&\quad-\{\lambda\|\hath-\hat h_{\lambda,t-1}^\text{IF}\|_2^2-\lambda\|\starh-\hat h_{\lambda,t-1}^\text{IF}\|_2^2\}\\
&\le(\E-\E_n)[\ell(V;\hath)-\ell(V;\starh)]\\
&\quad+\{\lambda\|\hath-h^*_{\lambda,t-1}\|_2^2-\lambda\|\starh-h^*_{\lambda,t-1}\|_2^2\}\\
&\quad-\{\lambda\|\hath-\hat h_{\lambda,t-1}^\text{IF}\|_2^2-\lambda\|\starh-\hat h_{\lambda,t-1}^\text{IF}\|_2^2\}\\
&\le(\E-\E_n)[\ell(V;\hath)-\ell(V;\starh)]\\
&\quad+2\lambda\|\hat h_{\lambda,t-1}^\text{IF}-h^*_{\lambda,t-1}\|_2\|\hath-\starh\|_2,
\end{align*}
\endgroup
and
\begin{equation}
\label{eq:thm6J2}	
\begin{aligned}
(\text{J}2)
&=
\|T\hath-\hat T\hath\|_2^2-\|T\starh-\hat T\starh\|_2^2\\
&~~~+2\E\left[\left\{(T\hath)(V_q)-(\hat T\hath)(V_q)\right\}\left\{\hat r(V_q)-r_0(V_q)\right\}\right]\\
&~~~-2\E\left[\left\{(T\starh)(V_q)-(\hat T\starh)(V_q)\right\}\left\{\hat r(V_q)-r_0(V_q)\right\}\right]\\
&\lesssim 
\|T-\hat T\|^2\|\hath-\starh\|_2
+\|T-\hat T\|\|\hat r-r\|_2\|\hath-\starh\|_2.
\end{aligned}
\end{equation}

With an approach similar to that in the proof of Theorem \ref{thm:mainNN0}, we have
\begingroup
\allowdisplaybreaks
\begin{align*}
&(\E-\E_n)[\ell(V;\hath)-\ell(V;\starh)]\\
&=(\E-\E_n)[\{(\hat T\hath)(V_q)-g_0(V)\}^2-\{(\hat T\starh)(V_q)-g_0(V)\}^2]\\
&\qquad+(\E-\E_n)[2\{(\hat T\hath)(V_q)-(\hat T\starh)(V_q)+(\hat T\starh)(V_q)-(\hat Th_0)(V_q)\\
&\qquad\qquad+(\hat Th_0)(V_q)-(Th_0)(V_q)+(Th_0)(V_q)-\hat r(V_q)\}g_1(V)\hath(V_h)]\\
&\qquad+(\E-\E_n)[-2\{(\hat T\starh)(V_q)-(\hat Th_0)(V_q)+(\hat Th_0)(V_q)-(Th_0)(V_q)+(Th_0)(V_q)-\hat r(V_q)\}g_1(V)\starh(V_h)]\\
&\qquad+(\E-\E_n)[-2\{(\hat T\hath)(V_q)-(\hat T\starh)(V_q)+(\hat T\starh)(V_q)-(\hat Th_0)(V_q)\\
&\qquad\qquad+(\hat Th_0)(V_q)-(Th_0)(V_q)+(Th_0)(V_q)-\hat r(V_q)\}(\hat T\hath)(V_q)]\\
&\qquad+(\E-\E_n)[2\{(\hat T\starh)(V_q)-(\hat Th_0)(V_q)+(\hat Th_0)(V_q)-(Th_0)(V_q)+(Th_0)(V_q)-\hat r(V_q)\}(\hat T\starh)(V_q)]\\
&\qquad+\lambda(\E-\E_n)[\{\hath(V_h)-\hat h_{\lambda,t-1}^\text{IF}(V_h)\}^2-\{\starh(V_h)-\hat h_{\lambda,t-1}^\text{IF}(V_h)\}^2]\\
&=(\E-\E_n)[\{(\hat T\hath)(V_q)-(\hat T\starh)(V_q)\}\{(\hat T\hath)(V_q)+(\hat T\starh)(V_q)-2g_0(V)\}]\\
&\qquad+(\E-\E_n)[2\{(\hat T\hath)(V_q)-(\hat T\starh)(V_q)\}\{g_1(V)\hath(V_h)-(\hat T\hath)(V_q)\}]\\
&\qquad+(\E-\E_n)[2\{(\hat T\starh)(V_q)-(\hat Th_0)(V_q)\}\{g_1(V)\hath(V_h)-g_1(V)\starh(V_h)+(\hat T\starh)(V_q)-(\hat T\hath)(V_q)\}]\\
&\qquad+(\E-\E_n)[2\{(\hat Th_0)(V_q)-(Th_0)(V_q)\}\{g_1(V)\hath(V_h)-g_1(V)\starh(V_h)-(\hat T\hath)(V_q)+(\hat T\starh)(V_q)\}]\\
&\qquad+(\E-\E_n)[2\{r_0(V_q)-\hat r(V_q)\}\{g_1(V)\hath(V_h)-g_1(V)\starh(V_h)-(\hat T\hath)(V_q)+(\hat T\starh)(V_q)\}]\\
&\qquad+\lambda(\E-\E_n)[\{\hath(V_h)-\starh(V_h)\}\{\hath(V_h)+\starh(V_h)-2\hat h_{\lambda,t-1}^\text{IF}(V_h)]\\
&\lesssim|(\E-\E_n)[(\hat T\hath)(V_q)-(\hat T\starh)(V_q)]|\\
&\qquad+|(\E-\E_n)[(\hat T\starh)(V_q)-(\hat Th_0)(V_q)]|\\
&\qquad+|(\E-\E_n)[(\hat Th_0)(V_q)-(Th_0)(V_q)]|\\
&\qquad+|(\E-\E_n)[r_0(V_q)-\hat r(V_q)]|\\
&\qquad+\lambda|(\E-\E_n)[\hath(V_h)-\starh(V_h)]|\\
&\lesssim\delta_{M,n}\|\hat T(\hath-\starh)\|_2+\delta_{M,n}^2\\
&\qquad+\delta_{M,n}\|\hat T(\starh-h_0)\|_2+\delta_{M,n}^2\\
&\qquad+\delta_{M,n}\|(\hat T-T)h_0\|_2+\delta_{M,n}^2\\
&\qquad+\delta_{M,n}\|\hat r-r_0\|_2+\delta_{M,n}^2\\
&\qquad+\lambda\delta_{n}\|\hath-\starh\|_2+\lambda\delta_{n}^2,
\end{align*}
\endgroup
with probability at least $1-5\zeta$, where the last inequality is due to Lemma \ref{lem:cons}.

Therefore, with probability at least $1-5\zeta$, we have
\begingroup
\allowdisplaybreaks
\begin{align*}
&(\text{J}1)\\
&\lesssim
\delta_{M,n}\{\|\hat T(\hath-\starh)\|_2+\|\hat T(\starh-h_0)\|_2+\|(\hat T-T)h_0\|_2+\|\hat r-r_0\|_2\}\\
&\qquad+\delta_{M,n}^2+\lambda\delta_{n}\|\hath-\starh\|_2+\lambda\delta_{n}^2+2\lambda\|\hat h_{\lambda,t-1}^\text{IF}-h^*_{\lambda,t-1}\|_2\|\hath-\starh\|_2\\
&\lesssim
\delta_{M,n}\{\|\hat T(\hath-\starh)\|_2+\|\hat T(\starh-h_0)\|_2+\|\hat T-T\|+\|\hat r-r_0\|_2\}\\
&\qquad+\delta_{M,n}^2+\lambda\delta_{n}^2+2\lambda\|\hat h_{\lambda,t-1}^\text{IF}-h^*_{\lambda,t-1}\|_2\|\hath-\starh\|_2.
\end{align*}
\endgroup
Note that
\begingroup
\allowdisplaybreaks
\begin{align*}
\|\hat T(\hath-\starh)\|_2
&\le \|(\hat T-T)(\hath-\starh)\|_2+\|T(\hath-\starh)\|_2\\
&\le \|\hat T-T\|+\|T(\hath-\starh)\|_2.
\end{align*}
\endgroup
Similarly,
\begin{align*}
\|\hat T(\starh-h_0)\|_2
\le \|\hat T-T\|+\|T(\starh-h_0)\|_2.
\end{align*}
Therefore, with probability at least $1-5\zeta$, we have
\begin{equation}
\label{eq:thm6J1}
\begin{aligned}
(\text{J}1)
&\lesssim
\delta_{M,n}\{\|T(\hath-\starh)\|_2+\|T(\starh-h_0)\|_2+\|\hat T-T\|+\|\hat r-r_0\|_2\}\\
&\qquad+\delta_{M,n}^2+\lambda\delta_{n}^2+2\lambda\|\hat h_{\lambda,t-1}^\text{IF}-h^*_{\lambda,t-1}\|_2\|\hath-\starh\|_2.
\end{aligned}
\end{equation}

Combining \eqref{eq:thm6J2} and \eqref{eq:thm6J1}, we conclude that, with probability at least $1-5\zeta$, we have
\begingroup
\allowdisplaybreaks
\begin{align*}
&\lambda\|\hath-\starh\|_2^2+\|T(\hath-\starh)\|_2^2\\
&\lesssim
\delta_{M,n}\{\|T(\hath-\starh)\|_2+\|T(\starh-h_0)\|_2+\|\hat T-T\|+\|\hat r-r_0\|_2\}\\
&\qquad+\|T-\hat T\|^2\|\hath-\starh\|_2
+\|T-\hat T\|\|\hat r-r\|_2\|\hath-\starh\|_2\\
&\qquad+\delta_{M,n}^2+\lambda\delta_{n}^2+2\lambda\|\hat h_{\lambda,t-1}^\text{IF}-h^*_{\lambda,t-1}\|_2\|\hath-\starh\|_2.
\end{align*}
\endgroup
Using the weighted AM-GM inequality,
\[
ab\le \frac{a^2}{2w}+\frac{wb^2}{2},
\]
we have
\begingroup
\allowdisplaybreaks
\begin{align*}
&\delta_{M,n}\{\|T(\hath-\starh)\|_2+\|T(\starh-h_0)\|_2+\|\hat T-T\|+\|\hat r-r_0\|_2\}\\
&\le \frac{1}{2w}\{\|T(\hath-\starh)\|_2+\|T(\starh-h_0)\|_2+\|\hat T-T\|+\|\hat r-r_0\|_2\}^2
+\frac{w}{2}\delta_{M,n}^2\\
&\le \frac{2}{w}\|T(\hath-\starh)\|_2^2
+\frac{2}{w}\|T(\starh-h_0)\|_2^2
+\frac{2}{w}\|\hat T-T\|^2+\frac{2}{w}\|\hat r-r_0\|_2^2
+\frac{w}{2}\delta_{M,n}^2.
\end{align*}
\endgroup
Therefore, there exists positive constants $c_1$ and $c_2$, with $c_1<1$, such that, with probability at least $1-5\zeta$, we have
\begingroup
\allowdisplaybreaks
\begin{align*}
\lambda\|\hath-\starh\|_2^2+\|T(\hath-\starh)\|_2^2
\lesssim~
&c_1\|T(\hath-\starh)\|_2^2\\
+&c_2\Big\{\|T(\starh-h_0)\|_2^2+\|\hat T-T\|^2+\|\hat r-r_0\|_2^2\\
&+\delta_{M,n}^2+\lambda\delta_{n}^2+2\lambda\|\hat h_{\lambda,t-1}^\text{IF}-h^*_{\lambda,t-1}\|_2\|\hath-\starh\|_2\Big\},
\end{align*}
\endgroup
which implies that, with probability at least $1-5\zeta$,
\begingroup
\allowdisplaybreaks
\begin{align*}
\lambda\|\hath-\starh\|_2^2+\|T(\hath-\starh)\|_2^2
&\lesssim
\|T(\starh-h_0)\|_2^2+\|\hat T-T\|^2+\|\hat r-r_0\|_2^2\\
&+\delta_{M,n}^2+\lambda\delta_{n}^2+2\lambda\|\hat h_{\lambda,t-1}^\text{IF}-h^*_{\lambda,t-1}\|_2\|\hath-\starh\|_2.
\end{align*}
\endgroup
By Lemma \ref{lem:reg} for projected error, under Assumption \ref{assm:beta}, we have
\begin{align*}
\|T(\starh-h_0)\|_2^2\le B\lambda^{\min\{2t,\beta+1\}}.
\end{align*}
Therefore, with probability at least $1-5\zeta$,
\begingroup
\allowdisplaybreaks
\begin{align*}
\lambda\|\hath-\starh\|_2^2+\|T(\hath-\starh)\|_2^2
&\lesssim
\|\hat T-T\|^2+\|\hat r-r_0\|_2^2+\delta_{M,n}^2\\
&+\lambda^{\min\{2t,\beta+1\}}+\lambda\delta_{n}^2+2\lambda\|\hat h_{\lambda,t-1}^\text{IF}-h^*_{\lambda,t-1}\|_2\|\hath-\starh\|_2.
\end{align*}
\endgroup
Next, we note that by the AM-GM inequality,
\begingroup
\allowdisplaybreaks
\begin{align*}
2\lambda\|\hat h_{\lambda,t-1}^\text{IF}-h^*_{\lambda,t-1}\|_2\|\hath-\starh\|_2
\le 
w\lambda\|\hat h_{\lambda,t-1}^\text{IF}-h^*_{\lambda,t-1}\|_2^2+
\frac{\lambda}{w}\|\hath-\starh\|_2^2
\end{align*}
\endgroup
Therefore, with a similar argument as before, there exists a choice of $w$ leading to 
\begingroup
\allowdisplaybreaks
\begin{align*}
\lambda\|\hath-\starh\|_2^2+\|T(\hath-\starh)\|_2^2
&\lesssim
\|\hat T-T\|^2+\|\hat r-r_0\|_2^2+\delta_{M,n}^2\\
&\quad+\lambda^{\min\{2t,\beta+1\}}+\lambda\delta_{n}^2+\lambda\|\hat h_{\lambda,t-1}^\text{IF}-h^*_{\lambda,t-1}\|_2^2.
\end{align*}
\endgroup
Recall that $\hat h_{\lambda,0}^\text{IF}=h^*_{\lambda,0}=0$.
Therefore, for $t=1$, we have
\begin{align*}
\lambda\|\hat h_{\lambda,1}^\text{IF}-h^*_{\lambda,1}\|_2^2
\lesssim 
\|\hat T-T\|^2+\|\hat r-r_0\|_2^2+\delta_{M,n}^2
+\lambda^{\min\{2,\beta+1\}}+\lambda\delta_{n}^2.	
\end{align*}
Therefore, for $\lambda<1$, with probability at least $1-5\zeta$,
\begin{align*}
\lambda\|\hat h_{\lambda,2}^\text{IF}-h^*_{\lambda,2}\|_2^2
+\|T(\hat h_{\lambda,2}^\text{IF}-h^*_{\lambda,2})\|_2^2
&\lesssim 
\|\hat T-T\|^2+\|\hat r-r_0\|_2^2+\delta_{M,n}^2+\lambda^{\min\{2,\beta+1\}}+\lambda\delta_{n}^2.
\end{align*}
This implies that, with probability at least $1-5\zeta$,
\begin{align*}
&\|\hat h_{\lambda,2}^\text{IF}-h^*_{\lambda,2}\|_2^2
\lesssim 
\frac{1}{\lambda}\big\{\|\hat T-T\|^2+\|\hat r-r_0\|_2^2+\delta_{M,n}^2\big\}+\lambda^{\min\{1,\beta\}}+\delta_{n}^2,\\
&\|T(\hat h_{\lambda,2}^\text{IF}-h^*_{\lambda,2})\|_2^2
\lesssim 
\|\hat T-T\|^2+\|\hat r-r_0\|_2^2+\delta_{M,n}^2+\lambda^{\min\{2,\beta+1\}}+\lambda\delta_{n}^2.
\end{align*}
Adding the regularization bias similar to the proof of Theorem \ref{thm:mainNN} concludes the desired result.

\begin{flushright}
$\Box$	
\end{flushright}

\subsection*{Proof of Theorem \ref{thm:hpc}}

For any given function $h$ and nuisance parameter pair $\tilde\eta=(\tilde T,\tilde r)$, we define
\begin{align*}
\ell(h,\tilde \eta)
&:=\left\{(\tilde Th)(V_q)-g_0(V)\right\}^2+2\left\{(\tilde Th)(V_q)- \tilde r(V_q)\right\}\left\{g_1(V)h(V_h)-(\tilde Th)(V_q)\right\}.
\end{align*}
Note that for $\eta_0=(T,r_0)$, i.e., the true values of the nuisance components, we have
\begin{align*}
\E[\ell(h,\eta_0)]
&=\E\left[\left\{(Th)(V_q)-g_0(V)\right\}^2+2\left\{(Th)(V_q)- r_0(V_q)\right\}\left\{g_1(V)h(V_h)-(Th)(V_q)\right\}\right]\\
&=\psi(h)=\E\left[\left\{(Th)(V_q)-g_0(V)\right\}^2\right].
\end{align*}
Moreover, our previously defined estimator $\hat\psi(h)$ can be written as
\begin{align*}
\hat\psi(h)
&=\E_n\left[\left\{(\hat Th)(V_q)-g_0(V)\right\}^2+2\left\{(\hat Th)(V_q)-\hat r(V_q)\right\}\left\{g_1(V)h(V_h)-(\hat Th)(V_q)\right\}\right]\\
&=\E_n[\ell(h,\hat\eta)].
\end{align*}

Based on this notation, we have
\[
h_0\in\argmin_{h\in L^2(P_{0,V_h})}\E[\ell(h,\eta_0)],
\]
and
\[
\hat h=\argmin_{h\in\mathcal{C}}\E_n[\ell(h,\hat\eta)].
\]
We first target bounding $\E[\ell(\hat h,\eta_0)]-\E[\ell(h^*,\eta_0)]$.

Define $L(h,\eta):=\ell(h,\eta)-\ell(h_0,\eta)$. Note that $\E[L(h,\eta_0)]\ge 0$, for all $h\in L^2(P_{0,V_h})$, and $\E[\ell(\hat h,\eta_0)]-\E[\ell(h^*,\eta_0)]=\E[L(\hat h,\eta_0)]-\E[L(h^*,\eta_0)]$. Therefore, we focus on the right hand side. This quantity can be decomposed as
\begingroup
\allowdisplaybreaks
\begin{align*}
&\E[L(\hat h,\eta_0)]-\E[L(h^*,\eta_0)]\\
&=\E_n[L(\hat h,\eta_0)]-\E_n[L(h^*,\eta_0)]\\
&\quad-(\E_n-\E)[L(\hat h,\eta_0)]\\
&\quad+(\E_n-\E)[L(h^*,\eta_0)]\\
&=\E_n[L(\hat h,\hat\eta)]-\E_n[L(h^*,\hat\eta)]\\
&\quad-(\E_n-\E)[L(\hat h,\eta_0)]\\
&\quad+(\E_n-\E)[L(h^*,\eta_0)]\\
&\quad-\E_n[L(\hat h,\hat\eta)-L(\hat h,\eta_0)]\\
&\quad+\E_n[L(h^*,\hat\eta)-L(h^*,\eta_0)]\\
&\le-(\E_n-\E)[L(\hat h,\eta_0)]\\
&\quad+(\E_n-\E)[L(h^*,\eta_0)]\\
&\quad-\E_n[L(\hat h,\hat\eta)-L(\hat h,\eta_0)]\\
&\quad+\E_n[L(h^*,\hat\eta)-L(h^*,\eta_0)],
\end{align*}
\endgroup
where the last inequality holds because $\E_n[L(\hat h,\hat\eta)]-\E_n[L(h^*,\hat\eta)]\le0$ as $h^*\in\mathcal{C}$. 
Following the ideas in \citep{vaart2006oracle,van2003unified}, for any $\delta>0$, the last expression is equal to
\begingroup
\allowdisplaybreaks
\begin{align*}
&-(1+\delta)(\E_n-\E)[L(\hat h,\eta_0)]-\delta\E[L(\hat h,\eta_0)]\\
&\quad+(1+\delta)(\E_n-\E)[L(h^*,\eta_0)]-\delta\E[L(h^*,\eta_0)]\\
&\quad-(1+\delta)\E_n[L(\hat h,\hat\eta)-L(\hat h,\eta_0)]\\
&\quad+(1+\delta)\E_n[L(h^*,\hat\eta)-L(h^*,\eta_0)]\\
&\quad+\delta\{\E_n[L(\hat h,\hat\eta)]-\E_n[L(h^*,\hat\eta)]\}\\
&\quad+2\delta\E[L(h^*,\eta_0)]\\
&\le-(1+\delta)(\E_n-\E)[L(\hat h,\eta_0)]-\delta\E[L(\hat h,\eta_0)]\\
&\quad+(1+\delta)(\E_n-\E)[L(h^*,\eta_0)]-\delta\E[L(h^*,\eta_0)]\\
&\quad-(1+\delta)\E_n[L(\hat h,\hat\eta)-L(\hat h,\eta_0)]\\
&\quad+(1+\delta)\E_n[L(h^*,\hat\eta)-L(h^*,\eta_0)]\\
&\quad+2\delta\E[L(h^*,\eta_0)],
\end{align*}
\endgroup
where the last inequality holds due to the same reason as above.

We write the last expression as 
\[
(T1)+(1+\delta)\{(T2)+(T3)\}+2\delta\E[L(h^*,\eta_0)],
\] 
where
\begin{align*}
(T1)
&=-(1+\delta)(\E_n-\E)[L(\hat h,\eta_0)]-\delta\E[L(\hat h,\eta_0)]\\
&\quad+(1+\delta)(\E_n-\E)[L(h^*,\eta_0)]-\delta\E[L(h^*,\eta_0)],\\
(T2)
&=-\E[L(\hat h,\hat\eta)-L(\hat h,\eta_0)]\\
&\quad+\E[L(h^*,\hat\eta)-L(h^*,\eta_0)],\\
(T3)
&=+(\E_n-\E)[L(h^*,\hat\eta)-L(h^*,\eta_0)]\\
&\quad-(\E_n-\E)[L(\hat h,\hat\eta)-L(\hat h,\eta_0)].
\end{align*}

We start with $(T2)$. We first note that
\begingroup
\allowdisplaybreaks
\begin{align*}
&\E[\ell(h,\hat\eta)-\ell(h,\eta_0)]\\
&=\E[\{(\hat Th)(V_q)-g_0(V)\}^2+2\{(\hat Th)(V_q)-\hat r(V_q)\}\{g_1(V)h(V_h)-(\hat Th)(V_q)\}\\
&\quad-\{( Th)(V_q)-g_0(V)\}^2-2\{( Th)(V_q)- r_0(V_q)\}\{g_1(V)h(V_h)-( Th)(V_q)\}]\\
&=\E[  ( Th)^2(V_q)-(\hat Th)^2(V_q)-2r_0(V_q)(\hat Th)(V_q)+2(Th)(V_q)(\hat Th)(V_q)-2\hat r(V_q)( Th)(V_q)\\
&\quad+2\hat r(V_q)(\hat Th)(V_q)-2( Th)^2(V_q)+2r_0(V_q)( Th)(V_q)-2r_0(V_q)( Th)(V_q)+2r_0(V_q)( Th)(V_q)]\\
&=\E[-\{( Th)(V_q)-(\hat Th)(V_q)\}^2+2\{( Th)(V_q)-(\hat Th)(V_q)\}\{r_0(V_q)-\hat r(V_q)\}].
\end{align*}
\endgroup
Therefore, 
\begingroup
\allowdisplaybreaks
\begin{align*}
(T2)
&=-\E[L(\hat h,\hat\eta)-L(\hat h,\eta_0)]\\
&\quad+\E[L(h^*,\hat\eta)-L(h^*,\eta_0)],\\
&=-\E[\ell(\hat h,\hat\eta)-\ell(h_0,\hat\eta)-\ell(\hat h,\eta_0)+\ell(h_0,\eta_0)]\\
&\quad+\E[\ell(h^*,\hat\eta)-\ell(h_0,\hat\eta)-\ell(h^*,\eta_0)+\ell(h_0,\eta_0)]\\
&=-\E[\ell(\hat h,\hat\eta)-\ell(\hat h,\eta_0)]\\
&\quad+\E[\ell(h^*,\hat\eta)-\ell(h^*,\eta_0)]\\
&=-\E[-\{( T\hat h)(V_q)-(\hat T\hat h)(V_q)\}^2+2\{( T\hat h)(V_q)-(\hat T\hat h)(V_q)\}\{r_0(V_q)-\hat r(V_q)\}]\\
&\quad+\E[-\{( Th^*)(V_q)-(\hat Th^*)(V_q)\}^2+2\{( Th^*)(V_q)-(\hat Th^*)(V_q)\}\{r_0(V_q)-\hat r(V_q)\}]\\
&=\E[\{( T\hat h)(V_q)-(\hat T\hat h)(V_q)\}^2-\{( Th^*)(V_q)-(\hat Th^*)(V_q)\}^2]\\
&\quad+2\E[\{( T\hat h)(V_q)-(\hat T\hat h)(V_q)-( Th^*)(V_q)+(\hat Th^*)(V_q)\}\{\hat r(V_q)-r_0(V_q)\}]\\
&=\E[\{( T\hat h)(V_q)-(\hat T\hat h)(V_q)-( Th^*)(V_q)+(\hat Th^*)(V_q)\}\\
&\quad\times\{( T\hat h)(V_q)-(\hat T\hat h)(V_q)+( Th^*)(V_q)-(\hat Th^*)(V_q)\}]\\
&\quad+2\E[\{( T\hat h)(V_q)-(\hat T\hat h)(V_q)-( Th^*)(V_q)+(\hat Th^*)(V_q)\}\{\hat r(V_q)-r_0(V_q)\}]\\
&=\E[((T-\hat T)(\hat h-h^*))(V_q)((T-\hat T)(\hat h+h^*))(V_q)]\\
&\quad+2\E[\{((T-\hat T)(\hat h-h^*))(V_q)\}\{\hat r(V_q)-r_0(V_q)\}]\\
&\le \|(T-\hat T)(\hat h-h^*)\|_2\|(T-\hat T)(\hat h+h^*)\|_2+2\|(T-\hat T)(\hat h-h^*)\|_2\|r_0-\hat r\|_2\\
&\le\|T-\hat T\|^2\|\hat h-h^*\|_2\|\hat h-h_0+h^*-h_0+2h_0\|_2+\|T-\hat T\|\|r_0-\hat r\|_2\|\hat h-h^*\|_2\\
&\le
\|T-\hat T\|^2\|\hat h-h^*\|_2\|\hat h-h_0\|_2
+\|T-\hat T\|^2\|\hat h-h^*\|_2\|h^*-h_0\|_2
+\|T-\hat T\|^2\|\hat h-h^*\|_2\|2h_0\|_2\\
&\quad+\|T-\hat T\|\|r_0-\hat r\|_2\|\hat h-h^*\|_2\\
&\lesssim\|T-\hat T\|^2\|\hat h-h^*\|_2+\|T-\hat T\|\|r_0-\hat r\|_2\|\hat h-h^*\|_2.
\end{align*}
\endgroup
We conclude that
\[
(T2)\lesssim\max_{h\in\mathcal{C}}\| h-h^*\|_2\|T-\hat T\|^2+\max_{h\in\mathcal{C}}\| h-h^*\|_2\|T-\hat T\|\|r_0-\hat r\|_2.
\]

Moving on to $(T3)$, we first note that
\begingroup
\allowdisplaybreaks
\begin{align*}
(T3)
&=(\E_n-\E)[L(h^*,\hat\eta)-L(h^*,\eta_0)]\\
&\quad-(\E_n-\E)[L(\hat h,\hat\eta)-L(\hat h,\eta_0)]\\
&=(\E_n-\E)[\ell(h^*,\hat\eta)-\ell(h_0,\hat\eta)-\ell(h^*,\eta_0)+\ell(h_0,\eta_0)]\\
&\quad-(\E_n-\E)[\ell(\hat h,\hat\eta)-\ell(h_0,\hat\eta)-\ell(\hat h,\eta_0)+\ell(h_0,\eta_0)]\\
&=(\E_n-\E)[\ell(h^*,\hat\eta)-\ell(h^*,\eta_0)]\\
&\quad-(\E_n-\E)[\ell(\hat h,\hat\eta)-\ell(\hat h,\eta_0)]\\
&=\E_n[\ell(h^*,\hat\eta)-\ell(h^*,\eta_0)-\ell(\hat h,\hat\eta)+\ell(\hat h,\eta_0)-\E[\ell(h^*,\hat\eta)-\ell(h^*,\eta_0)-\ell(\hat h,\hat\eta)+\ell(\hat h,\eta_0)]].
\end{align*}
\endgroup
Define $Z_{h,i}:=\ell_i(h^*,\hat\eta)-\ell_i(h^*,\eta_0)-\ell_i( h,\hat\eta)+\ell_i( h,\eta_0)-\E[\ell(h^*,\hat\eta)-\ell(h^*,\eta_0)-\ell( h,\hat\eta)+\ell( h,\eta_0)]$, which is a mean-zero random variable. In addition, define
\begingroup
\allowdisplaybreaks
\begin{align*}
T_n
&:=\max_{h\in\mathcal{C}}|\E_n[Z_{h,i}]|\\
&=\max_{h\in\mathcal{C}}|\frac{1}{n}\sum_{i=1}^nZ_{h,i}|.
\end{align*}
\endgroup
Our goal is to find a bound on $T_n$, satisfied with high probability, which will be a bound on $(T3)$, with high probability.

Since $h$'s and nuisance functions are estimated on separate data folds, $\sum_{i=1}^nZ_{h,i}$ is a sum of i.i.d. mean-zero random variables. 
Moreover, we have $\|Z_{h,1}\|_\infty<C$ for some constant $C$, and also, we have
\begingroup
\allowdisplaybreaks
\begin{align*}
var(Z_{h,1})
&\le \E[(\ell(h^*,\hat\eta)-\ell(h^*,\eta_0)-\ell( h,\hat\eta)+\ell( h,\eta_0))^2].
\end{align*}
\endgroup
Note that we have
\begingroup
\allowdisplaybreaks
\begin{align*}
&\E[(\ell(h^*,\hat\eta)-\ell(h^*,\eta_0)-\ell( h,\hat\eta)+\ell( h,\eta_0))^2]\\
&=\E[(\{(\hat Th^*)(V_q)-g_0(V)\}^2+2\{(\hat Th^*)(V_q)-\hat r(V_q)\}\{g_1(V)h^*(V_h)-(\hat Th^*)(V_q)\}\\
&\quad-\{( Th^*)(V_q)-g_0(V)\}^2-2\{( Th^*)(V_q)- r_0(V_q)\}\{g_1(V)h^*(V_h)-( Th^*)(V_q)\}\\
&\quad-\{(\hat Th)(V_q)-g_0(V)\}^2+2\{(\hat Th)(V_q)-\hat r(V_q)\}\{g_1(V)h(V_h)-(\hat Th)(V_q)\}\\
&\quad+\{( Th)(V_q)-g_0(V)\}^2-2\{( Th)(V_q)- r_0(V_q)\}\{g_1(V)h(V_h)-( Th)(V_q)\})^2]\\
&=\E[(( Th^*)^2(V_q)-(\hat Th^*)^2(V_q)+2g_0(V)\{( Th^*)(V_q)-(\hat Th^*)(V_q)\}+2g_1(V)h^*(V_h)\{(\hat Th^*)(V_q)-(Th^*)(V_q)\}\\
&\quad+2g_1(V)h^*(V_h)\{r_0(V_q)-\hat r(V_q)\}+2\hat r(V_q)(\hat Th^*)(V_q)-2r_0(V_q)(Th^*)(V_q)\\
&\quad-( T h)^2(V_q)+(\hat T h)^2(V_q)-2g_0(V)\{( T h)(V_q)-(\hat T h)(V_q)\}-2g_1(V) h(V_h)\{(\hat T h)(V_q)-(T h)(V_q)\}\\
&\quad-2g_1(V) h(V_h)\{r_0(V_q)-\hat r(V_q)\}-2\hat r(V_q)(\hat T h)(V_q)+2r_0(V_q)(T h)(V_q)
)^2]\\
&=\E[(( Th^*)^2(V_q)-(\hat Th^*)^2(V_q)-( T h)^2(V_q)+(\hat T h)^2(V_q)\\
&\quad+2g_0(V)\{( Th^*)(V_q)-(\hat Th^*)(V_q)\}-2g_0(V)\{( T h)(V_q)-(\hat T h)(V_q)\}\\
&\quad+2g_1(V)h^*(V_h)\{(\hat Th^*)(V_q)-(Th^*)(V_q)\}-2g_1(V) h(V_h)\{(\hat T h)(V_q)-(T h)(V_q)\}\\
&\quad+2g_1(V)h^*(V_h)\{r_0(V_q)-\hat r(V_q)\}-2g_1(V) h(V_h)\{r_0(V_q)-\hat r(V_q)\}\\
&\quad+2\hat r(V_q)(\hat Th^*)(V_q)-2r_0(V_q)(Th^*)(V_q)-2\hat r(V_q)(\hat T h)(V_q)+2r_0(V_q)(T h)(V_q)
)^2].
\end{align*}
\endgroup
We address each line in the last expression separately. The first line can be written as
\begingroup
\allowdisplaybreaks
\begin{align*}
&( Th^*)^2(V_q)-(\hat Th^*)^2(V_q)-( T h)^2(V_q)+(\hat T h)^2(V_q)\\
&=((T-\hat T)h^*)(V_q)((T+\hat T)h^*)(V_q)-((T-\hat T)h)(V_q)((T+\hat T)h)(V_q)\\
&\quad+((T-\hat T)h^*)(V_q)((T+\hat T)h)(V_q)-((T-\hat T)h^*)(V_q)((T+\hat T)h)(V_q)\\
&=((T-\hat T)h^*)(V_q)((T+\hat T)(h^*-h))(V_q)
+((T-\hat T)(h^*-h))(V_q)((T+\hat T)h)(V_q)\\
&=((T-\hat T)h^*)(V_q)((\hat T-T)(h^*-h))(V_q)
+2((T-\hat T)h^*)(V_q)(T(h^*-h))(V_q)\\
&\quad+((T-\hat T)(h^*-h))(V_q)((\hat T-T)h)(V_q)
+2((T-\hat T)(h^*-h))(V_q)(Th)(V_q)\\
&=((T-\hat T)(h^*-h_0))(V_q)((\hat T-T)(h^*-h))(V_q)
+((T-\hat T)(h_0))(V_q)((\hat T-T)(h^*-h))(V_q)\\
&\quad+2((T-\hat T)(h^*-h_0))(V_q)(T(h^*-h))(V_q)
+2((T-\hat T)(h_0)(V_q)(T(h^*-h))(V_q)\\
&\quad+((T-\hat T)(h^*-h))(V_q)((\hat T-T)(h-h_0))(V_q)
+((T-\hat T)(h^*-h))(V_q)((\hat T-T)h_0)(V_q)\\
&\quad+2((T-\hat T)(h^*-h))(V_q)(T(h-h_0))(V_q)
+2((T-\hat T)(h^*-h))(V_q)(Th_0)(V_q).
\end{align*}
\endgroup
The second line can be written as
\begingroup
\allowdisplaybreaks
\begin{align*}
&2g_0(V)\{( Th^*)(V_q)-(\hat Th^*)(V_q)\}-2g_0(V)\{( T h)(V_q)-(\hat T h)(V_q)\}\\
&=2g_0(V)((T-\hat T)(h^*-h))(V_q).
\end{align*}
\endgroup
The third line can be written as
\begingroup
\allowdisplaybreaks
\begin{align*}
&2g_1(V)h^*(V_h)\{(\hat Th^*)(V_q)-(Th^*)(V_q)\}-2g_1(V) h(V_h)\{(\hat T h)(V_q)-(T h)(V_q)\}\\
&=2g_1(V)h^*(V_h)((\hat T-T)h^*)(V_q)-2g_1(V)h(V_h)((\hat T-T)h)(V_q)\\
&\quad+2g_1(V)h(V_h)((\hat T-T)h^*)(V_q)-2g_1(V)h(V_h)((\hat T-T)h^*)(V_q)\\
&=2g_1(V)\{h^*(V_h)-h(V_h)\}((\hat T-T)h^*)(V_q)
+2g_1(V)h(V_h)((\hat T-T)(h^*-h))(V_q)\\
&=2g_1(V)\{h^*(V_h)-h(V_h)\}((\hat T-T)(h^*-h_0))(V_q)
+2g_1(V)\{h^*(V_h)-h(V_h)\}((\hat T-T)h_0)(V_q)\\
&\quad+2g_1(V)(h-h_0)(V_h)((\hat T-T)(h^*-h))(V_q)
+2g_1(V)h_0(V_h)((\hat T-T)(h^*-h))(V_q).
\end{align*}
\endgroup
The fourth line can be written as
\begingroup
\allowdisplaybreaks
\begin{align*}
&2g_1(V)h^*(V_h)\{r_0(V_q)-\hat r(V_q)\}-2g_1(V) h(V_h)\{r_0(V_q)-\hat r(V_q)\}\\
&=2g_1(V)\{h^*(V_h)-h(V_h)\}\{r_0(V_q)-\hat r(V_q)\}
\end{align*}
\endgroup
Finally, the fifth line can be written as
\begingroup
\allowdisplaybreaks
\begin{align*}
&2\hat r(V_q)(\hat Th^*)(V_q)-2r_0(V_q)(Th^*)(V_q)-2\hat r(V_q)(\hat T h)(V_q)+2r_0(V_q)(T h)(V_q)\\
&=2\hat r(V_q)(\hat T(h^*-h))(V_q)-2r_0(V_q)(T(h^*-h))(V_q)
+2 r_0(V_q)(\hat T(h^*-h))(V_q)-2 r_0(V_q)(\hat T(h^*-h))(V_q)\\
&=2\{\hat r(V_q)-r_0(V_q)\}(\hat T(h^*-h))(V_q)
+2 r_0(V_q)((\hat T-T)(h^*-h))(V_q)\\
&=2\{\hat r(V_q)-r_0(V_q)\}((\hat T-T)(h^*-h))(V_q)
+2\{\hat r(V_q)-r_0(V_q)\}(T(h^*-h))(V_q)\\
&\quad+2 r_0(V_q)((\hat T-T)(h^*-h))(V_q).
\end{align*}
\endgroup
Therefore, we have
\begingroup
\allowdisplaybreaks
\begin{align*}
&\E[(\ell(h^*,\hat\eta)-\ell(h^*,\eta_0)-\ell( h,\hat\eta)+\ell( h,\eta_0))^2]\\
&\lesssim\|T-\hat T\|^2\|h-h^*\|_2^2+\|r_0-\hat r\|^2_2\| h-h^*\|^2_2\\
&\lesssim\max_{h\in\mathcal{C}}\|h-h^*\|_2^2\|T-\hat T\|^2+\max_{h\in\mathcal{C}}\| h-h^*\|^2_2\|r_0-\hat r\|^2_2.
\end{align*}
\endgroup
Hence, we have an upper bound $\sigma^2=c(\max_{h\in\mathcal{C}}\|h-h^*\|_2^2\|T-\hat T\|^2+\max_{h\in\mathcal{C}}\| h-h^*\|^2_2\|r_0-\hat r\|^2_2)$ on $var(Z_{h,1})$, for some constant $c$.

Therefore, by Lemma \ref{lem:Bern}, for all $x>0$, we have
\begingroup
\allowdisplaybreaks
\begin{align*}
\E[T_n>\frac{x}{n}]
&=\E[\max_{h\in\mathcal{C}}|\frac{1}{n}\sum_{i=1}^nZ_{h,i}|>\frac{x}{n}]\\
&=\E[\bigcup_{h\in\mathcal{C}}\{|\frac{1}{n}\sum_{i=1}^nZ_{h,i}|>\frac{x}{n}\}]\\
&\le\sum_{h\in\mathcal{C}}\E[|\frac{1}{n}\sum_{i=1}^nZ_{h,i}|>\frac{x}{n}]\\
&\le\underbrace{2M\exp\{\frac{-1}{2}\frac{x^2}{n\sigma^2+Cx}\}}_{\zeta}.
\end{align*}
\endgroup
Therefore,
\begingroup
\allowdisplaybreaks
\begin{align*}
&2\underbrace{\log(2M/\zeta)}_{b}=\frac{x^2}{n\sigma^2+Cx}\\
&\Rightarrow x^2-2Cbx-2bn\sigma^2=0\\
&\Rightarrow x=Cb\pm \sqrt{C^2b^2+2bn\sigma^2}.
\end{align*}
\endgroup
But $x>0$, hence $x=Cb+ \sqrt{C^2b^2+2bn\sigma^2}$.
This implies that with probability at least $1-\zeta$,
\[
T_n\le\frac{x}{n}=\frac{1}{n}\{Cb+ \sqrt{C^2b^2+2bn\sigma^2}\}.
\]

Recall that $\sqrt{a+b}\le\sqrt{a}+\sqrt{b}$, for $a,b\ge0$. Therefore, we have
\begingroup
\allowdisplaybreaks
\begin{align*}
T_n
&\le\frac{2Cb}{n}+\frac{\sqrt{2b}\sigma}{\sqrt{n}}\\
&\lesssim\frac{\log (2M/\zeta)}{n}+
\sqrt{\log(2M/\zeta)}\frac{1}{\sqrt{n}}\sqrt{\max_{h\in\mathcal{C}}\|h-h^*\|_2^2\|T-\hat T\|^2+\max_{h\in\mathcal{C}}\| h-h^*\|^2_2\|r_0-\hat r\|^2_2}\\
&\le \frac{\log (2M/\zeta)}{n}+
\sqrt{\log(2M/\zeta)}\frac{1}{\sqrt{n}}\left\{\max_{h\in\mathcal{C}}\|h-h^*\|_2\|T-\hat T\|+\max_{h\in\mathcal{C}}\| h-h^*\|_2\|r_0-\hat r\|_2\right\}.
\end{align*}
\endgroup

We conclude that with probability at least $1-\zeta$,
\[
(T3)\lesssim\frac{\log (2M/\zeta)}{n}+
\sqrt{\log(2M/\zeta)}\frac{1}{\sqrt{n}}\left\{\max_{h\in\mathcal{C}}\|h-h^*\|_2\|T-\hat T\|+\max_{h\in\mathcal{C}}\| h-h^*\|_2\|r_0-\hat r\|_2\right\}.
\]

Moving on to $(T1)$, we first note that
\begingroup
\allowdisplaybreaks
\begin{align*}
&(\E_n-\E)[L( h,\eta_0)]\\
&=\E_n[L_i( h,\eta_0)-\E[L( h,\eta_0)]].
\end{align*}
\endgroup
Define $Z_{h,i}:=L_i( h,\eta_0)-\E[L( h,\eta_0)]$, which is a mean-zero random variable. In addition, define
\begingroup
\allowdisplaybreaks
\begin{align*}
T_n
&:=\max_{h\in\mathcal{C}}(1+\delta)|\E_n[Z_{h,i}]|-\delta\E[L(h,\eta_0)]\\
&=\max_{h\in\mathcal{C}}(1+\delta)|\frac{1}{n}\sum_{i=1}^nZ_{h,i}|-\delta\E[L(h,\eta_0)].
\end{align*}
\endgroup
Our goal is to find a bound on $T_n$, satisfied with high probability, which then implies that $(T3)\le2(\text{bound on }T_n)$ with high probability.

Since $h$'s and nuisance functions are estimated on separate data folds, $\sum_{i=1}^nZ_{h,i}$ is a sum of i.i.d. mean-zero random variables. Moreover, we have $\|Z_{h,1}\|_\infty<C$ for some constant $C$, and also, we have
\begingroup
\allowdisplaybreaks
\begin{align*}
var(Z_{h,1})
&\le \E[\{L( h,\eta_0)\}^2]\\
&= \E[\{\ell( h,\eta_0)-\ell( h_0,\eta_0)\}^2].
\end{align*}
\endgroup

Note that we have
\begingroup
\allowdisplaybreaks
\begin{align*}
&\E[\{\ell( h,\eta_0)-\ell( h_0,\eta_0)\}^2]\\
&=\E[(\{(Th)(V_q)-g_0(V)\}^2+2\{(Th)(V_q)-r_0(V_q)\}\{g_1(V)h(V_h)-(Th)(V_q)\}\\
&\quad-\{( Th_0)(V_q)-g_0(V)\}^2-2\{(Th_0)(V_q)-r_0(V_q)\}\{g_1(V)h_0(V_h)-(Th_0)(V_q)\})^2]\\
&=\E[(( Th_0)^2(V_q)-(Th)^2(V_q)+2g_0(V)\{(Th_0)(V_q)-(Th)(V_q)\}\\
&\quad+2r_0(V_q)\{g_1(V)h_0(V_h)-g_1(V)h(V_h)\}+2r_0(V_q)\{(Th)(V_q)-(Th_0)(V_q)\}\\
&\quad+2g_1(V)h(V_h)\{(Th)(V_q)-(Th_0)(V_q)\}
+2(Th_0)(V_q)\{g_1(V)h(V_h)-g_1(V)h_0(V_h)\}
)^2]\\
&=\E[(\{(Th_0)(V_q)-(Th)(V_q)\}\{(Th_0)(V_q)+(Th)(V_q)\}
+2g_0(V)\{(Th_0)(V_q)-(Th)(V_q)\}\\
&\quad+2r_0(V_q)\{(Th)(V_q)-(Th_0)(V_q)\}+2g_1(V)h(V_h)\{(Th)(V_q)-(Th_0)(V_q)\}\\
&\quad
+2\{(Th_0)(V_q)-2r_0(V_q)\}\{g_1(V)h(V_h)-g_1(V)h_0(V_h)\}
)^2]\\
&=\E[(\{(Th_0)(V_q)-(Th)(V_q)\}\{(Th_0)(V_q)+(Th)(V_q)\}
+2g_0(V)\{(Th_0)(V_q)-(Th)(V_q)\}\\
&\quad+2r_0(V_q)\{(Th)(V_q)-(Th_0)(V_q)\}+2g_1(V)h(V_h)\{(Th)(V_q)-(Th_0)(V_q)\}\
)^2]\\
&\lesssim\|T(h-h_0)\|_2^2.
\end{align*}
\endgroup
Hence, we have an upper bound $\sigma^2=c\|T(h-h_0)\|_2^2$ on $var(Z_{h,1})$, for some constant $c$.

Therefore, by Lemma \ref{lem:Bern}, for all $x>0$, we have
\begingroup
\allowdisplaybreaks
\begin{align*}
&\E\left[(1+\delta)\left|\frac{1}{n}\sum_{i=1}^nZ_{h,i}\right|-\delta\E[L(h,\eta_0)]>x\right]\\
&=\E\left[\left|\frac{1}{n}\sum_{i=1}^nZ_{h,i}\right|>\frac{1}{1+\delta}x+\frac{\delta}{1+\delta}\E[L(h,\eta_0)]\right]\\
&\le2\exp\left\{\frac{-1}{2}\frac{n^2\left\{\frac{1}{1+\delta}x+\frac{\delta}{1+\delta}\E[L(h,\eta_0)]\right\}^2}{n\sigma^2+{nC}\left\{\frac{1}{1+\delta}x+\frac{\delta}{1+\delta}\E[L(h,\eta_0)]\right\}}\right\}\\
&=2\exp\left\{\frac{-1}{2}\frac{\left\{\frac{1}{1+\delta}x+\frac{\delta}{1+\delta}\E[L(h,\eta_0)]\right\}^2}{\frac{\sigma^2}{n}+\frac{C}{n}\left\{\frac{1}{1+\delta}x+\frac{\delta}{1+\delta}\E[L(h,\eta_0)]\right\}}\right\}.
\end{align*}
\endgroup

Therefore,
\begingroup
\allowdisplaybreaks
\begin{align*}
&\E \left[T_n>x\right]\\
&=\E\left[\max_{h\in\mathcal{C}}(1+\delta)\left|\frac{1}{n}\sum_{i=1}^nZ_{h,i}\right|-\delta\E[L(h,\eta_0)]>x\right]\\
&=\E\left[\bigcup_{h\in\mathcal{C}}\left\{(1+\delta)\left|\frac{1}{n}\sum_{i=1}^nZ_{h,i}\right|-\delta\E[L(h,\eta_0)]>x\right\}\right]\\
&\le\sum_{h\in\mathcal{C}}\E\left[(1+\delta)\left|\frac{1}{n}\sum_{i=1}^nZ_{h,i}\right|-\delta\E[L(h,\eta_0)]>x\right]\\
&\le\sum_{h\in\mathcal{C}}2\exp\left\{\frac{-1}{2}\frac{\left\{\frac{1}{1+\delta}x+\frac{\delta}{1+\delta}\E[L(h,\eta_0)]\right\}^2}{\frac{\sigma^2}{n}+\frac{C}{n}\left\{\frac{1}{1+\delta}x+\frac{\delta}{1+\delta}\E[L(h,\eta_0)]\right\}}\right\}\\
&\le M\max_{h\in\mathcal{C}}2\exp\left\{\frac{-1}{2}\frac{\left\{\frac{1}{1+\delta}x+\frac{\delta}{1+\delta}\E[L(h,\eta_0)]\right\}^2}{\frac{\sigma^2}{n}+\frac{C}{n}\left\{\frac{1}{1+\delta}x+\frac{\delta}{1+\delta}\E[L(h,\eta_0)]\right\}}\right\}.
\end{align*}
\endgroup
Suppose the maximum above is achieved at $\tilde h$.
As noted in the appendix of \citep{vaart2006oracle},
\begin{align*}
\frac{(y+\lambda)^2}{\frac{\sigma^2}{n}+\frac{C}{n}(x+\lambda)}\ge 
\begin{cases}
	\frac{y\lambda+\lambda^2}{2\frac{\sigma^2}{n}}\qquad &\text{ if }y\le\frac{\sigma^2}{C}-\lambda,\\
	~\\
	\frac{y+\lambda}{2\frac{C}{n}}\qquad &\text{ if }y\ge\frac{\sigma^2}{C}-\lambda,
\end{cases}
\end{align*}
where $y=\frac{1}{1+\delta}x$, and $\lambda=\frac{\delta}{1+\delta}\E[L(\tilde h,\eta_0)]$. Therefore,
\begingroup
\allowdisplaybreaks
\begin{align*}
&\E \left[T_n>x\right]\\
&\le2M\exp\left\{\frac{-1}{2}\frac{y\lambda+\lambda^2}{2\frac{\sigma^2}{n}}\right\}I\left(y\le\frac{\sigma^2}{C}-\lambda\right)\\
&\quad+2M\exp\left\{\frac{-1}{2}\frac{y+\lambda}{2\frac{C}{n}}\right\}I\left(y\ge\frac{\sigma^2}{C}-\lambda\right).
\end{align*}
\endgroup

With $\zeta=2M\exp\left\{\frac{-1}{2}\frac{y\lambda+\lambda^2}{2\frac{\sigma^2}{n}}\right\}$, we have $x=\frac{1}{n}\frac{1}{1+\delta}\frac{\sigma^2}{\lambda}4\log\frac{2M}{\zeta}-\lambda$,
and with $\zeta=2M\exp\left\{\frac{-1}{2}\frac{y+\lambda}{2\frac{C}{n}}\right\}$, we have $x=\frac{1}{n}\frac{1}{1+\delta}C4\log\frac{2M}{\zeta}-\lambda$.
Therefore, with probability at least $1-\zeta$, we have
\begingroup
\allowdisplaybreaks
\begin{align*}
T_n
&\le \frac{1}{n}\frac{1}{1+\delta}\left\{\frac{\sigma^2}{\lambda}+C\right\}4\log\frac{2M}{\zeta}-\lambda\\
&\le\frac{1}{n}\left\{\frac{\sigma^2}{\lambda}+C\right\}4\log\frac{2M}{\zeta}.
\end{align*}
\endgroup
where $\lambda=\frac{\delta}{1+\delta}\E[L(\tilde h,\eta_0)]$.

Suppose we can show $\frac{\sigma^2}{\lambda}\le C'$, for some constant $C'$. Then we can conclude that with probability at least $1-2\zeta$, we have
\[
(T1)\lesssim\frac{\log(2M/\zeta)}{n}.
\]
Therefore, it remains to prove the boundedness of $\frac{\sigma^2}{\lambda}$.

Note that
\begin{align*}
\frac{\sigma^2}{\lambda}
&= \frac{1+\delta}{\delta}\frac{\sigma^2}{\E[L(\tilde h,\eta_0)]}\\
&\asymp\frac{1+\delta}{\delta}\frac{\|T(\tilde h-h_0)\|_2^2}{\E[L(\tilde h,\eta_0)]}.
\end{align*}
Set $\delta=1$. Regarding $\E[L(\tilde h,\eta_0)]$, note that
\begingroup
\allowdisplaybreaks
\begin{align}
&\E[L(\tilde h,\eta_0)] \notag \\
&=\E[\ell(\tilde h,\eta_0)-\ell( h_0,\eta_0)] \notag \\
&=\E[\{(T\tilde h)(V_q)-g_0(V)\}^2+2\{(T\tilde h)(V_q)-r_0(V_q)\}\{g_1(V)\tilde h(V_h)-(T\tilde h)(V_q)\} \notag \\
&\quad-\{( Th_0)(V_q)-g_0(V)\}^2-2\{(Th_0)(V_q)-r_0(V_q)\}\{g_1(V)h_0(V_h)-(Th_0)(V_q)\}] \notag \\
&=\E[\{(T\tilde h)(V_q)-g_0(V)\}^2-\{( Th_0)(V_q)-g_0(V)\}^2] \notag \\
&=\E[(T\tilde h)^2(V_q)+r_0^2(V_q)-2r_0(V_q)(T\tilde h)(V_q)-(T h_0)^2(V_q)-r_0^2(V_q)+2r_0(V_q)(Th_0)(V_q)] \notag \\
&=\E[\{(T\tilde h)(V_q)-r_0(V_q)\}^2-\{( Th_0)(V_q)-r_0(V_q)\}^2] \notag \\
&=\E[\{(T\tilde h)(V_q)-r_0(V_q)\}^2] \notag \\
&=\E[\{(T\tilde h)(V_q)-( Th_0)(V_q)\}^2] \notag \\
&=\|T(\tilde h-h_0)\|_2^2.\label{eq:TisE}
\end{align}
\endgroup

Therefore,
\[
\frac{\sigma^2}{\lambda}\lesssim 2\frac{\|T(\tilde h-h_0)\|_2^2}{\|T(\tilde h-h_0)\|_2^2}=2.
\]

~\\

From the obtained bounds for $(T1)$, $(T2)$, and $(T3)$, we conclude that with probability at least $1-3\zeta$,
\begingroup
\allowdisplaybreaks
\begin{equation}
	\label{eq:ptg}
\begin{aligned}
&\E[L(\hat h,\eta_0)]-\E[L(h^*,\eta_0)]\\
&\lesssim 2\E[L(h^*,\eta_0)]
+\frac{\log (2M/\zeta)}{n}\\
&\quad+\sqrt{\log(2M/\zeta)}\frac{1}{\sqrt{n}}\left\{\max_{h\in\mathcal{C}}\|h-h^*\|_2\|T-\hat T\|+\max_{h\in\mathcal{C}}\| h-h^*\|_2\|r_0-\hat r\|_2\right\}\\
&\quad+\max_{h\in\mathcal{C}}\| h-h^*\|_2\|T-\hat T\|^2+\max_{h\in\mathcal{C}}\| h-h^*\|_2\|T-\hat T\|\|r_0-\hat r\|_2\\
&\lesssim 2\E[L(h^*,\eta_0)]
+\frac{\log (2M/\zeta)}{n}\\
&\quad+\max_{h\in\mathcal{C}}\| h-h^*\|_2\|T-\hat T\|^2+\max_{h\in\mathcal{C}}\| h-h^*\|_2^2\|\hat r-r_0\|_2^2+\max_{h\in\mathcal{C}}\| h-h^*\|_2\|T-\hat T\|\|\hat r-r_0\|_2.
\end{aligned}
\end{equation}
\endgroup

As we saw in Display \eqref{eq:TisE}, $\|T(h-h_0)\|_2^2=\E[L(h,\eta_0)]$. Therefore, using inequality \eqref{eq:ptg}, with probability at least $1-3\zeta$, we have
\begingroup
\allowdisplaybreaks
\begin{align*}
\|T(\hat h-h_0)\|_2^2
&=\|T( h^*-h_0)\|_2^2\\
&\quad+\|T( \hat h-h_0)\|_2^2-\|T( h^*-h_0)\|_2^2\\
&=\|T( h^*-h_0)\|_2^2\\
&\quad+\E[L(\hat h,\eta_0)]-\E[L(h^*,\eta_0)]\\
&\lesssim \|T( h^*-h_0)\|_2^2\\
&\quad+2\|T( h^*-h_0)\|_2^2
+
\frac{\log (2M/\zeta)}{n}\\
&\quad+\max_{h\in\mathcal{C}}\| h-h^*\|_2\|T-\hat T\|^2+\max_{h\in\mathcal{C}}\| h-h^*\|_2^2\|\hat r-r_0\|_2^2\\
&\quad+\max_{h\in\mathcal{C}}\| h-h^*\|_2\|T-\hat T\|\|\hat r-r_0\|_2,
\end{align*}
\endgroup	
which implies that
\begin{align*}
\big\|T(\hat h-h_0)\big\|_2^2
&\lesssim
\big\|T(h^*-h_0)\big\|_2^2\\
&\quad+\frac{\log (2M/\zeta)}{n}+\max_{h\in\mathcal{C}}\| h-h^*\|_2\|T-\hat T\|^2
+\max_{h\in\mathcal{C}}\| h-h^*\|_2^2\|\hat r-r_0\|_2^2.
\end{align*}

\begin{flushright}
$\Box$	
\end{flushright}

\subsection*{Proof of Corollary \ref{cor:thm}}

Based on the result of Theorem \ref{thm:hpc}, it suffices to bound $\big\|T(h^*-h_0)\big\|_2^2$.

Let $\lambda^*_1$ be a choice of the hyper-parameter stated in Theorem \ref{thm:mainNN}, and let $\lambda_\mathcal{C}^*$ be the smallest element in $\{\lambda_i\}_{i=1}^n$ which is larger than or equal to $\lambda^*_1$. That is,
\[
\lambda^*_1\le\lambda_\mathcal{C}^*\le\lambda^*_1+B_n/n.
\]
Moreover, let $h^*_1:=\hat h_{\lambda^*_\mathcal{C},2}^\text{IF}$, that is,
$h^*_1$ corresponds to hyper parameter $\lambda_\mathcal{C}^*$.

Hence, using Theorem \ref{thm:mainNN}, with probability at least $1-2\zeta$, we have
\begingroup
\allowdisplaybreaks
\begin{equation}
\label{eq:cor2}
\begin{aligned}
\|T(h^*-h_0)\|_2^2
&\le
\|T(h^*_1-h_0)\|_2^2\\
&\lesssim
(\lambda^*_1+B_n/n)^{\min\{4,\beta+1\}}\\
&\lesssim
\left(\Delta_n^{\frac{1}{\min\{5,\beta+2\}}}+\frac{B_n}{n}\right)^{\min\{4,\beta+1\}}\\
&\le
\left(2\Delta_n^{\frac{1}{\min\{5,\beta+2\}}}\right)^{\min\{4,\beta+1\}}\\
&\lesssim\Delta_n^{\frac{\min\{4,\beta+1\}}{\min\{5,\beta+2\}}}.
\end{aligned}
\end{equation}
\endgroup

Let $\lambda^*_2$ be a choice of the hyper-parameter among the candidates satisfying $\lambda^*_2\lesssim\Delta_{M,n}^{\frac{1}{\min\{2,\beta+1\}}}$.
Let $h^*_2$ be the corresponding candidate function.

Hence, using Theorem \ref{thm:mainNN1}, with probability at least $1-5\zeta$, we have
\begin{equation}
\label{eq:cor1}
\begin{aligned}
\|T(h^*-h_0)\|_2^2
&\le
\|T(h^*_2-h_0)\|_2^2\\
&\lesssim
\Delta_{M,n}+\delta_{n}^2.
\end{aligned}
\end{equation}

Therefore, \eqref{eq:cor2} and \eqref{eq:cor1} conclude that
\begin{align*}
\big\|T(h^*-h_0)\big\|_2^2
&\lesssim
\min\Big\{
\Delta_n^{\frac{\min\{4,\beta+1\}}{\min\{5,\beta+2\}}},\Delta_{M,n}+\delta_{n}^2
\Big\}.
\end{align*}

\begin{flushright}
$\Box$	
\end{flushright}

\subsection*{Proof of Theorem \ref{thm:hpcs}}

Regarding the source error, using Jensen's inequality similar to the approach of \citep{florens2011identification}, we have
\begingroup
\allowdisplaybreaks
\begin{align*}
&\|\hat h-h_0\|_2^{2(1+\alpha)}\\
&=\left(\sum_{i=1}^\infty\langle h-h_0,\phi_i\rangle^2\right)^{1+\alpha}\\
&=\left(\frac{\sum_{i=1}^\infty\langle h-h_0,\phi_i\rangle^2}{\sum_{i=1}^\infty\frac{\langle h-h_0,\phi_i\rangle^2}{\mu_i}}\sum_{i=1}^\infty\frac{\langle h-h_0,\phi_i\rangle^2}{\mu_i}\right)^{1+\alpha}\\
&=\left(\frac{\sum_{i=1}^\infty\langle h-h_0,\phi_i\rangle^2}{\sum_{i=1}^\infty\frac{\langle h-h_0,\phi_i\rangle^2}{\mu_i}}\right)^{1+\alpha}\left(\sum_{i=1}^\infty\frac{\langle h-h_0,\phi_i\rangle^2}{\mu_i}\right)^{1+\alpha}\\
&=\left(\sum_{i=1}^\infty\mu_i\frac{\frac{\langle h-h_0,\phi_i\rangle^2}{\mu_i}}{\sum_{j=1}^\infty\frac{\langle h-h_0,\phi_j\rangle^2}{\mu_j}}\right)^{1+\alpha}\left(\sum_{i=1}^\infty\frac{\langle h-h_0,\phi_i\rangle^2}{\mu_i}\right)^{1+\alpha}\\
&\le\left(\sum_{i=1}^\infty\mu_i^{1+\alpha}\frac{\frac{\langle h-h_0,\phi_i\rangle^2}{\mu_i}}{\sum_{j=1}^\infty\frac{\langle h-h_0,\phi_j\rangle^2}{\mu_j}}\right)\left(\sum_{i=1}^\infty\frac{\langle h-h_0,\phi_i\rangle^2}{\mu_i}\right)^{1+\alpha}\\
&=\left(\frac{\sum_{i=1}^\infty\mu_i^{\alpha}\langle h-h_0,\phi_i\rangle^2}{\sum_{j=1}^\infty\frac{\langle h-h_0,\phi_j\rangle^2}{\mu_j}}\right)\left(\sum_{i=1}^\infty\frac{\langle h-h_0,\phi_i\rangle^2}{\mu_i}\right)^{1+\alpha}\\
&=\left(\sum_{i=1}^\infty\mu_i^{\alpha}\langle h-h_0,\phi_i\rangle^2\right)\left(\sum_{i=1}^\infty\frac{\langle h-h_0,\phi_i\rangle^2}{\mu_i}\right)^{\alpha}\\
&\overset{(a)}{\lesssim}\left(\sum_{i=1}^\infty\mu_i^{\alpha}\langle h-h_0,\phi_i\rangle^2\right)\\
&\overset{(b)}{\le}\|T(\hat h-h_0)\|_2^2,
\end{align*}
\endgroup	
where $(a)$ and $(b)$ follow from Parts 1 and 2 of Assumption \ref{assm:alpha}, respectively. Hence, we conclude that
\[
\|\hat h-h_0\|_2^{2}\lesssim\|T(\hat h-h_0)\|_2^{\frac{2}{1+\alpha}}.
\]

\begin{flushright}
$\Box$	
\end{flushright}

\section{Auxiliary Lemmas}

\begin{lemma}[Bound on the regularization bias]
\label{lem:reg}
Consider the function $\starh$ defined in Equation \eqref{eq:popobjNN}. Under Assumption \ref{assm:beta}, the regularization bias can be bounded as
\begin{equation}
\label{eq:mpf22}
\|\starh-h_0\|_2^2\le B\lambda^{\min\{2t,\beta\}},
\hspace{10mm}
\|T(\starh-h_0)\|_2^2\le B\lambda^{\min\{2t,\beta+1\}}.
\end{equation}
\end{lemma}
See \citep{carrasco2007linear} for a proof for the case of compact operators, and \citep{engl1996regularization} for a proof for the general case. See also \citep{bennett2023source}.

\begin{lemma}[Localized concentration inequality \citep{foster2019orthogonal}]
\label{lem:cons}
Let $\Hc$ be a uniformly bounded function class, i.e., $\sup_{h\in\Hc}\|h\|_\infty$ is bounded by a constant.
Let $\ell$ be a real-valued loss function that is $L$-Lipschitz in its argument $h$ with respect to $L^2(P_0)$-norm.
Let $h^*$ be any fixed element of $\Hc$, and let $\delta_n$ be an upper bound on the critical radius of $star(\Hc-h^*)$ which satisfies $\delta_n^2\ge c_1\left( \frac{\log(\log(n))+\log(1/\zeta)}{n} \right)$, for some constant $c_1$.
Then, for some constant $c_2$, with probability at least $1-\zeta$, we have
\[
\big|(\E-\E_n)\left[\ell(V;h)-\ell(V;h^*)\right]\big|\le c_2\left(\delta_n\|h-h^*\|_2+\delta_n^2\right),~~~~~\forall h\in\Hc.
\]
\end{lemma}

\begin{lemma}[Bernstein inequality \citep{vaart2023empirical}]
\label{lem:Bern}
Let $Y_1,...,Y_n$ be i.i.d. mean-zero random variables. Let $var(Y_1)=v$ and $\|Y_1\|_\infty/3=M$. For all $x>0$, we have
\[
\E\left[\left|\frac{1}{n}\sum_{i=1}^nY_i\right|>\frac{x}{n}\right]\le2\exp\left\{-\frac{1}{2}\frac{x^2}{nv+Mx}\right\}.
\]
\end{lemma}

\end{document}